%% file: Kinsler-HEMW-2022arxiv.tex
\documentclass[aps,pra,a4paper,10pt,twocolumn,nofootinbib]{revtex4-2} 
\usepackage[T1]{fontenc}
\usepackage{mathptmx}
\usepackage{datetime}
\usepackage{amsmath}
\usepackage{amsfonts}
\usepackage{amsfonts}
\usepackage{mathrsfs}
\usepackage[mathscr]{euscript}
\usepackage[dvips]{graphicx}
\usepackage{colordvi}
\usepackage{color}
\usepackage{hyperref} 
\usepackage{epsfig}
\usepackage{bm}
\usepackage{enumitem}
\usepackage{graphics,color}

\def\overstrike#1#2{{\setbox0\hbox{$#2$}\hbox to \wd0{\hss
    $#1$\hss}\kern-\wd0\box0}}

\newdateformat{yymmdddate}{\THEYEAR/\twodigit{\THEMONTH}/\twodigit{\THEDAY}}

\begin{document}
\title{Agent swarms: cooperation and coordination under stringent communications constraint}

\author{Paul Kinsler}
\homepage[]{https://orcid.org/0000-0001-5744-8146}
\email{Dr.Paul.Kinsler@physics.org}
\affiliation{
  Department of Electronic \& Electrical Engineering
  University of Bath, Bath, BA2 7AY,
  United Kingdom
}

\author{Sean Holman}
\homepage[]{https://orcid.org/0000-0001-8050-2585}
\affiliation{Department of Mathematics, University of Manchester, Manchester, M13 9PL, United Kingdom}

\author{Andrew Elliott}
\homepage[]{https://orcid.org/0000-0002-4536-5244}
\affiliation{School of Mathematics and Statistics, University of Glasgow, Glasgow, G12 8QQ, United Kingdom}

\author{Cathryn N. Mitchell}
\homepage[]{https://orcid.org/0000-0003-1964-8723}
\affiliation{
  Department of Electronic and Electrical Engineering
  University of Bath, Bath, BA2 7AY,
  United Kingdom
}

\author{R. Eddie Wilson} 
\affiliation{Department of Engineering Mathematics, University of Bristol, Bristol, BS8 1TW, United Kingdom}

\begin{abstract}

Here we consider the communications tactics 
 appropriate 
 for a group of agents
 that need to ``swarm'' together 
 in a highly adversarial environment.
Specifically, 
 whilst they need to cooperate by
 exchanging information with each other 
 about their location and their plans; 
 at the same time they also need to keep 
 such communications to an absolute minimum.
This might be due to 
 a need for stealth, 
 or otherwise be relevant to situations where 
 communications are significantly restricted.
Complicating this process
 is that we assume each agent has 
 (a) no means
 of passively locating others,
 (b) it must rely on being updated by reception of appropriate messages;
 and if no such update messages arrive,
 (c) then their own beliefs about other agents
 will gradually become out of date and increasingly inaccurate.
Here we use a geometry-free multi-agent model 
 that is capable of allowing for message-based information transfer
 between agents with different intrinsic connectivities, 
 as would be present in a spatial arrangement of agents.
We present agent-centric performance metrics that 
 require only minimal assumptions, 
 and show how simulated outcome distributions, 
 risks, 
 and connectivities
 depend on the ratio of information gain to loss.
We also show that 
 checking for too-long round-trip-times
 can be an effective minimal-information filter 
 for determining 
 which agents to no longer target with messages.
\end{abstract}


\date{\today}
\maketitle

\def\overstrike#1#2{{\setbox0\hbox{$#2$}\hbox to \wd0{\hss
    $#1$\hss}\kern-\wd0\box0}}




\def\tinyzero{{\scalebox{0.6}{0}}}

\def\Aknow{\Phi}
\def\Akmin{\Aknow_{\textup{m}}}
\def\Elink{L}

\def\Trate{\alpha}
\def\Trsum{A}
\def\Trmax{\bar{\Trsum}}
\def\Rrisk{R}

\def\inforate{q}

\def\upT{\textup{T}}
\def\pthreshold{{\Aknow}_{\upT}}

\def\nticks{K}

%
\section{Introduction}\label{S-intro}

For more than a decade, 
 the availability and range of applications for Unmanned Aerial Vehicles (UAVs)
 or ``drones'' 
 has greatly increased.
{They now span areas such as
 logistics, agriculture, remote sensing,
 communication,
 security,
 and defense
 \cite{Bekmezcia-ST-2013ahn,Shakhatreh-SFDAKOKG-2019ieeea,Haider-NJJBK-2022mdpi,Rejeb-ART-2022cea,Benarbia-K-2022mdpi,Kucharczyk-H-2021rse}.
In particular, 
 UAVs appear to be useful
 for tasks which are too dangerous, expensive, 
 or innaccessible by manned vehicles. 
They also offer the advantage 
 of being able to use a swarm of smaller and less expensive drones
 in place of a single UAV.
In addition, 
 the operating capabilities may differ across individual elements for the swarm. 
Groups of drones operating cooperatively have also been proposed for surveillance
 \cite{Stodola-NFVTBPNS-2022mesa,Shakhatreh-SFDAKOKG-2019ieeea,Maza-CCMO-2011jirs},
 search and rescue \cite{Cameron-SW-2010suaave,Waharte-TJ-2009suaave,Horyna-BWAHFS-2023ar},
 and military missions \cite{Grimal-S-2019jcsl,George-SS-2011jirs}.
Using a swarm has the advantage
 of being robust against losses of individual drones.}
Further, 
 since smaller and more agile individual swarm constituents
 can be difficult to detect and attack,
 they can have 
 a natural advantage for applications in which the environment is contested.

Combining a group of drones into a ``swarm"
 is a difficult engineering problem
 with potential to draw from a wide range of disciplines
 \cite{Stodola-NFVTBPNS-2022mesa,Horyna-BWAHFS-2023ar,Grimal-S-2019jcsl,Saeed-OAAA-2022}.
{The advantage is that it
 acts and can be directed as a single entity
 while minimising communication
 and remaining robust to both physical and cyber attacks.
Swarm robotics and swarm engineering are emerging fields \cite{Brambilla-FBD-2013si,Chung-PDSK-2018tor,Roldan-CB-2018icirs,Saeed-OAAA-2022}
 which study automated decision making and control
 for large groups of robots using only local communication between nearby swarm members.
Much work in that field considers larger swarms ($\approx 10^2-10^5$)
 than of interest for flying drone swarms ($\approx 10-10^2)$. 
There are special challenges inherent in the creation and real-time maintenance
 of non-centralised ad hoc networks \cite{Horyna-BWAHFS-2023ar,Hayat-YM-2016cst,Arafat-M-2019itj}
 for groups of UAVs.
Note that these are
 variously termed FANETs (Flying Ad Hoc Networks)
 \cite{Bekmezcia-ST-2013ahn}
 and UAANETs (UAV Ad Hoc Networks) \cite{Maxa-ML-2017ahswm}; 
 these have been studied in recent years
 and various architectures and protocols
 have been proposed \cite{Haider-NJJBK-2022mdpi,Bekmezcia-ST-2013ahn,Maxa-ML-2017ahswm}. 
A number of groups have developed
 \textit{in silico} test-beds for FANETs \cite{Mairaja-BJ-2019smpt,Vasarhelyi-VSNEV-2018sr,Albani-MSF-2022icra}
 with some moving on to \textit{in robotico} realisations,
 and there has been recent interest in local algorithms
 for adaptive FANETs as well as consensus \cite{Chen-LG-2020tvt,Vasarhelyi-VSNEV-2018sr,Maza-CCMO-2011jirs}.}
{However, 
 work with specific application to search and rescue \cite{Cameron-SW-2010suaave,Waharte-TJ-2009suaave,Horyna-BWAHFS-2023ar},
 rarely considers network resilience to external attacks
 or the need to minimise risk of detection, 
 although
 there has been research on autonomous swarms with covert leaders \cite{Han-RS-2007robo}.}

%

In this work we focus on scenarios 
 in which the drones (hereafter \emph{agents})
 must act in the extreme limit of minimal information sharing.
This is most easily represented as situations where 
 communications must be restricted in order to maintain \emph{stealth}.
This is the opposite of typical scenarios, 
 where information sharing and other communications 
 are considered as ``free''.
In such a typical scenario,
 each agent almost automatically has an excellent knowledge
 as to the state of the swarm, 
 or at least some coordinating agent or supervisor has such information
 with which to efficiently direct swarm operations.
It should be noted that 
 this ``cooperation under communications constraint''
 is a different problem to \emph{control} under communications constraint
 \cite{Jain-SV-2003cdc,Tatikonda-M-2004tac}.

In the minimal information cases we consider in this paper 
 we should no longer
 talk of what an agent ``knows'', 
 since that unhelpfully suggests a likely misleading degree of accuracy;
 but instead what it ``believes'', 
 since that implies an appropriate degree of doubt and uncertainty.
Although up to date and accurate information might be 
 received in any new message just received, 
 since communications will be sparse, 
 we need to keep in mind that this will nevertheless
 become less reliable as time passes.
What this means in practice is that many of the typical treatments
 of swarm activities as given above 
 (e.g. \cite{Haider-NJJBK-2022mdpi,Bekmezcia-ST-2013ahn,Rejeb-ART-2022cea,Benarbia-K-2022mdpi,Kucharczyk-H-2021rse,Stodola-NFVTBPNS-2022mesa,Horyna-BWAHFS-2023ar,Brambilla-FBD-2013si,Roldan-CB-2018icirs})
 become secondary problems
 to the fundamental issue \cite{CEME-DASA-2022}:
 i.e. how well does each agent know where the others are, 
 so that it might send a message, 
 and what tactics should it use to 
 maximise the accuracy of its beliefs, 
 whilst minimising its exposure to adversarial action?
A game theory \cite{GeckilAnderson-AGAME,Farooqui-N-2016casm} problem 
 arises here because 
 an agent gains no new information
 by sending a message, 
 but such transmissions only expose it to more risk.
Instead, 
 remaining silent might allow agents to risklessly accumulate information
 from the transmissions of others --
 except that if all agents do this,
 the swarm cannot behave coherently.
{The situation is
 a comparable}
 to an inverse tragedy of the commons \cite{Diekert-2012s}.

{In our scenario, 
 agents must both cooperate and coordinate.
Here ``cooperation'' refers to the necessity that all agents 
 must cooperate by all sending location messages, 
 because without this, 
 agents will end up with inaccurate information, 
 and so be unable to target messages successfully, 
 so that swarm connectivity must then fail.
Further, 
 ``coordination'' refers to the fact that a connected swarm
  can only be formed using multi-hop message paths
  (as per Sec. 4.3)
  if other agents cooperate by forwarding such 
  messages.
 Without this cooperation, 
  any message must be sent directly,  
  requiring accurate information about all other agents.
 In such a direct-signalling paradigm,
  not only would 
  an increased transmission power be needed to reach the longer ranges, 
  increasing risk, 
  but any blocked messaged path would 
  be fatal for connectivity.}

{Our contribution here is that we
 construct a mathematical multi-agent information model}
 that can be 
 specialized to both continuum communications
 and discrete communications models.
The discrete communications model is then 
 used to design a stochastic simulation code, 
 which we used to 
 benchmark and test a minimal approach that 
 optimises communications
 whilst minimising risk, 
 whilst using only minimal assumptions.
{In particular,
 we focus in on the underlying basics}
 of the ``to transmit ... or not to transmit?'' problem
 without the many additional complications of 
 as movement in space,
 specific communications physics,
 or how to build optimal communications networks 
 within the swarm.
{As well as considering what properties of the model
 an agent might be permitted to use when taking action, 
 we present some performance and risk metrics
 that help evaluate the performance-under-constraint scenario, 
 and also propose a round-trip timing test that 
 is based solely on data an agent is aware of,
 and 
 which can be used
 to deprecate poor links.}

{In what follows we present our 
 model and its concepts in Sec. \ref{S-geomfree}.
This is 
 followed by a continuum communication implementation
 in Sec. \ref{S-geomfree-continuum}, 
 where an agent might act
 to modify its transmission priorities
 on the basis of
 the \emph{rate} of information arrival 
 from other agents.
This rate-equation description can be used to generate indicative 
 steady-state answers,
 and allow some preliminary conclusions.
To assist with judgements about performance, 
 we define some agent and swarm metrics
 in Sec. \ref{S-geomfree-metrics}.
Then we describe our
 implementation of a discrete communications approach
 in Sec. \ref{S-geomfree-stochastic}, 
 where we 
 use a Monte-Carlo approach to produce 
 a more sophisticated understanding 
 of the distribution of possible outcomes.
Here, 
 the informational basis on which an agent might 
 act is no longer a rate, 
 but instead the timings (and time-delays) 
 of messages from other agents, 
 and so in Sec. \ref{S-estatic}
 we show and explain some results using this approach.
After a discussion in Sec. \ref{S-discussion}, 
 we conclude in Sec. \ref{S-conclusion}.}

%
\section{Multi-agent model}\label{S-geomfree}

{We consider a swarm of $N$ agents
 that intend to cooperate and send messages 
 about their activities to each other.}
Our model state is intended to mimic the behavior
 of a spatially distributed swarm of agents, 
 but without requiring a detailed spatial model
 and all the additional complications that would entail.
{As such it contains only 
 a minimal set of features,
 and is primarily intended to facilitate an initial understanding 
 of how our novel scenario might be handled.}
{The model contains 
 four types of information, 
 intended to represent as simply as possible 
 the accuracy of an agents beliefs,
 the degradation of that accuracy as time passes, 
 the environment's effect on signalling efficiency, 
 and agent messaging choices.
Thus:}

\begin{description}

\item[\emph{First},] 
 each agent $a$ has an information store 
 about all other agents $j$, 
 {which we summarize using values ${\Aknow}^a_j$.
If this store contains recent and reliable information, 
 we would expect it to result in 
 a high probability of messaging success, 
 but if the information is outdated or otherwise unreliable,  
 the probability would instead be low.}
Thus
 these accuracies are represented as probabilities, 
 using
 real numbers ${\Aknow}^a_j \in [0,1]$;
 with zero representing entirely inaccurate beliefs
 and 1 representing perfectly accurate beliefs.
Since agent $a$ is presumably perfectly informed about itself, 
 ${\Aknow}^a_a = 1$ should always hold.

\item[\emph{Second},]
 we allow for the possibility that 
 an agent's beliefs about others
 slowly become out of date and degraded.
We model this by assuming that all the ${\Aknow}^a_j$ 
 (iff $j \neq a$)
 decay exponentially as determined by some loss parameter $\gamma$.
However, 
 we assume there is also a minimum ``find by chance'' probability ${\Akmin}$ 
 such that ${\Aknow}^a_j \geq {\Akmin}$
 for any $a$ and all $j$.

\item[\emph{Third},]
 there is an agent-to-agent transmission efficiency, 
 which represents environmental constraints
 {that might hinder communications between agents.
This agent-to-agent transmission efficiency ${\Elink}_{ab} \in [0,1]$
 enables the representation of a wide range of networks, 
 including 
 networks based on spatial positions and signal models, 
 as well as 
 those with ${\Elink}_{ab}$ generated randomly according to some algorithm, 
 e.g. abstract Erdos-Renyi (ER) networks
 \cite{Coscia-NetworkAtlas}.
However, 
 at this early stage we do not specify how the efficiencies
 ${\Elink}_{ab}$ might have been generated, 
 and can even allow ${\Elink}_{ab} \neq {\Elink}_{ba}$, 
 i.e. the transmission efficiency from $a$ to $b$ 
 may be different to the transmission efficiency from $b$ to $a$.
An important feature is that 
 we do not assume any agent $a$ has any information 
 about either ${\Elink}_{ba}$ or ${\Elink}_{ab}$.}

\item[\emph{Fourth},]
 since an agent might transmit information 
 at different rates towards different targets, 
 we also specify its set of transmission rates ${\Trate}^a_i$.

\end{description}

\begin{figure}
\begin{center}
\resizebox{0.750\columnwidth}{!}{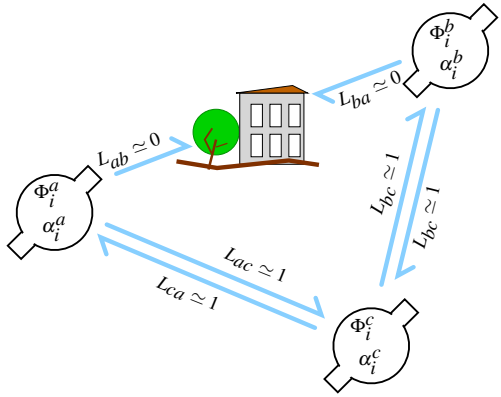}
\end{center}
\caption{Diagram of a simple swarm with three agent-drones
 $a$, $b$, and $c$; 
 where the link $a \leftrightarrow b$ is blocked
 so that ${\Elink}_{ab}$ and ${\Elink}_{ba}$ are near zero, 
 whereas the $a \leftrightarrow c$ and $b \leftrightarrow c$ 
 links are free of obstruction.
This means we would expect the values ${\Aknow}^a_b$ and ${\Aknow}^b_a$
 to be small (since the beliefs, being poorly updated,
 will become ever more
 inaccurate), 
 whereas the unobstructed messaging 
 along links $a \leftrightarrow c$ and $b \leftrightarrow c$
 should mean that it is possible to maintain 
 ${\Aknow}^a_c$, ${\Aknow}^c_a$, ${\Aknow}^b_c$, and ${\Aknow}^c_b$
 at values near 1. 
}
\label{fig-agentio}
\end{figure}

These definitions 
 mean that
 while the probability of message transmission
 from an agent $a$ to another $b$
 is straighforwardly given by the product of ${\Aknow}^a_b$ and ${\Elink}_{ab}$, 
 the actual rate of information arrival
 is ${\Trate}^a_i {\Aknow}^a_b {\Elink}_{ab}$.
This model is broadly consistent with an implict assumption
 that transmisions are sent directionally and \emph{need} to be aimed, 
 being sent from $a$ to $b$ 
 with an accuracy ${\Aknow}^a_b$
 across a link with efficiency ${\Elink}_{ab}$.
Also,
 in this model,
 messages are only ever received by the intended recipient.
{A simple depiction with just three agents
 and one blocked (inefficient) link 
 is given 
 in fig.\ref{fig-agentio}.}

%
\subsection{Index convention: subscruipts and superscripts}\label{S-geomfree-indices}

{To enable easier interpretation of the model parameters and values, 
 we use an index convention
 where each indexing letter, 
 and its positioning as a super- or sub-script, 
 implies extra meaning.
If we are referring to some 
 specific agent we use one of 
 $\{a, b, c\}$, 
 where $a,b,c \in \{1, 2, ..., N\}$; 
 but if referring to a range of other agents
 will use one of  
 $\{i, j, k\}$, 
 where $i,j,k = \{1, 2, ..., N\}$. 
Further, 
 a superscript denotes that the quantity 
 is a property of that superscripted agent, 
 but for a subscript,
 there is no such implication.
That is, 
 the value ${\Aknow}^a_j$ is a property of $a$ for any $j$, 
 but for none of those $j$ (if $a \ne j$) is it a property.
Thus ${\Aknow}^a_b$ is a number that is a property of agent $a$, 
 and
 ${\Aknow}^a_i$ is a collection of numbers that is a property of agent $a$.
However, 
 the collection of 
 ${\Aknow}^i_j$ (or even ${\Aknow}^i_a$) 
 is global information.
This is because 
 $i$ and $j$ each encompasses many agents, 
 so that ${\Aknow}^i_j$
 is not a property
 of any single agent in the swarm.}
Other characters used as sub- or superscripts 
 will indicate not agents but special cases or particular values
 of e.g. ${\Aknow}$ or ${\Elink}$.
We do \emph{not} use any implied summation convention.

%
\subsection{Abstractions are not knowledge}\label{S-geomfree-abstracto}

{When using models of the type proposed here, 
 it is important to note that 
 an agent property (e.g. ${\Aknow}^a_i$)
 is defined within the model
 as being attributable to
 an agent $a$, 
 and may affect the outcomes of $a$'s actions.
However, 
 even though the model of agent $a$
 contains a collection of properties, 
 this \emph{does not}
 mean that the agent decision making
 can necessarily make use of each and every agent property.
In particular, 
 here we have that 
 ${\Aknow}^a_i$ is a representation or 
 \emph{abstraction} of agent $a$'s beliefs about the spatial location of agent $i$, 
 thus telling the model how efficiently agent $a$ can target that agent $i$.
This is why the model will use it when calculating 
 either what fraction of the information contained
 in the messages sent actually arrives, 
 as in the continuum communications model; 
 or alternatively whether or not a whole message is received, 
 as in the discrete communications model.}

{Despite this,
 there is no reason why any \emph{actual} agent $a$
 will be aware of and be able to use the value of ${\Aknow}^a_i$
 in decision making.
For example,
 the model could specify that an agent $a$
 has an accuracy ${\Aknow}^a_b = 0.50$
 when messaging $b$.
However, 
 the agent $a$ might not be \emph{aware} that that is the accuracy, 
 so that 
 it cannot use its value of $0.50$ when decision making,
 e.g. by using it in a formula or algorithm.}
This is because an actual agent 
 will instead only be cognisant of 
 some specific ``basket'' of data --
 containing entries such as position estimates, 
 likely errors, 
 future plans for movement, 
 and so on --
 which need not be reducible in an algorithmic way 
 to the model's substitute, 
 i.e. the abstraction ${\Aknow}^a_i$'s particular value.
{That is, 
   any \emph{actual} agent $a$ might only be aware of a basket of specific details,
    but not how to synthesise the abstract ${\Aknow}^a_j$
    from those details.}
   Indeed, 
    when \emph{we} use this model, 
    we do not know this synthesising process either, 
    nor anything about the basket contents.

\begin{figure}
\begin{center}
\resizebox{0.750\columnwidth}{!}{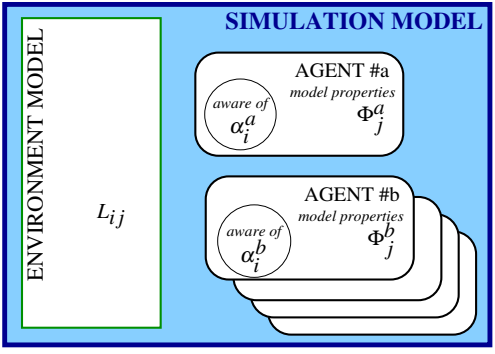}
\end{center}
\caption{{
Schematic showing model (or simulation) parameters, 
 i.e. 
 environment ($\Elink_{ij}$)
 and agent properties (${\Aknow}^b_a$, ${\Trate}^b_a$), 
 and indicating that
 an agent is only aware of 
 some of the agent model's properties
 (here, only ${\Trate}^b_a$).
Depending on the interpretation intended by a specific model, 
 the information loss $\gamma$ might 
 be in any one of these categories; 
 or it might even be a combination
 of both environment and agent properties.
}}
\label{fig-venndiagram}
\end{figure}

As discussed above, 
 {and as indcated in fig. \ref{fig-venndiagram},}
 we have that ${\Aknow}^a_i$ is a property of the agent $a$ model, 
 but that agent $a$ is not aware of ${\Aknow}^a_j$.
In contrast, 
 an agent should be aware of its
 choices or settings for transmission rates ${\Trate}^a_i$, 
 so these would be usable in decision making.
However, 
 whether parameters such as $\gamma$ or ${\Akmin}$
 {are agent properties that the agent is aware of, 
 agent properties that are unknown, 
 or even parameters entirely unrelated to the agent description,
 will depend on how we envisage the
 model of agent loss and message transmission.
Nevertheless, 
 since an agent $a$ is unaware of 
 the agent property ${\Aknow}^a_j$,
 it seems reasonable to decide likewise that $\gamma$
 is also (at best) only an (unknown) agent property. 
However, 
 if $\gamma$ were (e.g.) dependent on the environment, 
 it might not even be considered an agent property.}

This distinction between the model's abstractions
 and an agent's actual awareness or beliefs -- 
 whatever they might be -- 
 means that to
 implement a generalisable communications tactic
 we
 must avoid reliance on our model's abstractions, 
 and instead use only quantities
 that an agent is aware of, 
 can measure, 
 or believes.

%
\subsection{Terminology}\label{S-geomfree-terminology}

In this work we will often refer to a ``link'', 
 meaning the potential for communication 
 between two agents $a$ and $b$.
However,
 here ``link'' is just a short and convenient word 
 for that potential, 
 and it does not imply that such a communication 
 is guaranteed to be easy --
 or even possible --
 in any particular case.
{When discussing links, 
 we will also use three adjectives --
 efficent, accurate, reliable -- 
 to describe them, 
 and these three adjectives have specific meanings
 which we will now define.
However, 
 this terminology is only intended 
 to make general discussion clearer by specifying preferred adjectives,
 rather than as any unique mathematical specification; 
 and for descriptive purposes the exact threshold value
 is not of primary importance.}

For some link between two agents $a$ and $b$, 
 we say that there is an ``efficient'' link 
 if ${\Elink}_{ab}$ is sufficiently large, 
 i.e. if it exceeds some suitable threshold value, 
 as discussed later --
 e.g. perhaps if ${\Elink}_{ab} > 0.75$; 
 conversely it is an ``inefficient'' link 
 if it does not.

When communicating over these links, 
 the agents will send messages, 
 and depending on the model, 
 this may deliver information either gradually and incrementally, 
 as in our continuum communications model;
 or in packets, 
 as in our discrete communications model.
The primary purpose of these messages is that they enable
 accurate targetting of replies back to the sending agent,
 but in our discrete model they also contain timing 
 information as to when the last message on that link was received.

When considering targetting, 
 for some link between two agents $a$ and $b$, 
 we say that there is ``accurate'' targetting of messages
 from $a$ to $b$ if ${\Aknow}^a_b$ is sufficiently large, 
 e.g. if it exceeds some suitable threshold value ${\Aknow}_{\upT}$;
 conversely it is ``inaccurate'' 
 if it does not.

Lastly,
 for some link between two agents $a$ and $b$, 
 we say that there is a ``reliable'' link 
 from $a$ to $b$ if the product ${\Aknow}^a_b {\Elink}_{ab}$ is sufficiently large, 
 e.g. if it exceeds some suitable threshold value;
 conversely it is an ``unreliable'' link 
 if it does not.

%
\section{Continuum communications}\label{S-geomfree-continuum}

In this continuum model, 
 we aim to represent a system in which all agents
 are simultaneously feeding trickles of information 
 to all other agents,
 with no randomness or contingency in the process.
Although not very realistic, 
 this rate equation model
 gives us a good starting point
 {with which to introduce 
 parameters and concepts in a simple and direct way.}
In the latter part of this paper, 
 we will 
 move to a more plausible stochastic model
 based on discrete messaging choices.

First, 
 we assume that each agent $b$ 
 transmits to each other agent $a$, 
 sending information about itself (only)
 according to an information rate ${\Trate}^b_a$, 
 and with a targetting accuracy dependent on its imperfect ${\Aknow}^b_a$.
Further, 
 the information stream sent will be attenuated in transit 
 by the link efficiency ${\Elink}_{ba}$.

As a result 
 $a$ will only be able to improve its ${\Aknow}^a_b$
 at a maximum rate 
 $\inforate^a_b = {\Trate}^b_a {\Elink}_{ab} {\Aknow}^b_a$.    
Further, 
 when agent $a$ receives some information in a message, 
 it will only find the currently unknown part of that information
 useful.
Thus 
 only a fraction $1-{\Aknow}^a_b$ of that arriving from $b$
 will add to the existing total ${\Aknow}^a_b$.

Here we assume that this receive rate $\inforate^a_b$ can be measured
 therefore the receiving agent will be aware of it, 
 but that the agent cannot measure its constituent contributions
 ${\Trate}^b_a$,
 ${\Elink}_{ab}$, 
 and ${\Aknow}^b_a$ individually.
However, 
 we do allow that 
 the receiving agent might still be able to make plausible inferences
 about those unknown contributions
 by making some assumptions.

%
\subsection{Dynamics}\label{S-geomfree-continuum-dyn}

Based on the description and parameters given above, 
 we can now write a rate equation 
 for the behaviour of each ${\Aknow}^a_b$, 
 for $a \neq b$,
 which is
~
\begin{align}
  {\frac{d}{dt}}
  {\Aknow}^a_b
&=
 -
  \gamma 
  \left( {\Aknow}^a_b - {\Akmin} \right)
 +
  {\Trate}^b_a
  {\Elink}_{ba}
  {\Aknow}^b_a
   \left(1-{\Aknow}^a_b\right)
.
\label{eqn-continuum-rate}
\end{align}
In what follows we will typically assume that 
 the $\gamma$, $\Elink_{ba}$, ${\Trate}^b_a$, 
 and ${\Akmin}$ values are fixed parameters, 
 and only the ${\Aknow}^a_b$ are time-dependent\footnote{Although
   we do not do so here, 
   if we were to also to permit an agent $b$'s beliefs 
   about its own position to not be perfectly accurate, 
   i.e. if ${\Aknow}^b_b < 1$, 
   then the second term on the right hand side of 
   \eqref{eqn-continuum-rate}
   should also be multiplied by ${\Aknow}^b_b$; 
   since it is passing on imperfect information.
  However, 
   we might also need an additional or modified rate equation
   to determine how each of ${\Aknow}^i_i$ might behave; 
   or reinterpret ${\Trate}^a_a$ or ${\Elink}_{aa}$
   as providing a supply of new self-location information.}.

%
\subsection{Steady state}\label{S-geomfree-continuum-static}

Since there are no complicated interdependencies, 
 it is straightforward to get a steady-state solution
 by setting $(d/dt) {\Aknow}^a_b = 0$.
This gives
~
\begin{align}
  \gamma 
  \left( {\Aknow}^a_b - {\Akmin} \right)
&=
  {\Trate}^b_a
  {\Elink}_{ab}
  {\Aknow}^b_a
   \left(1-{\Aknow}^a_b\right)
\\
  \gamma {\Aknow}^a_b
 -
  \gamma {\Akmin}
&=
  {\Trate}^b_a
  {\Elink}_{ba}
  {\Aknow}^b_a
 -
  {\Trate}^b_a
  {\Elink}_{ba}
  {\Aknow}^b_a
  {\Aknow}^a_b
\\
  \left(
    \gamma 
   +
    {\Trate}^b_a
    {\Elink}_{ba}
    {\Aknow}^b_a
  \right)
  {\Aknow}^a_b
&=
  \gamma {\Akmin}
 +
  {\Trate}^b_a
  {\Elink}_{ba}
  {\Aknow}^b_a
\\
  {\Aknow}^a_b
&=
  \frac{  \gamma {\Akmin}  +  {\Trate}^b_a {\Elink}_{ba} {\Aknow}^b_a  }
       {  \gamma           +  {\Trate}^b_a {\Elink}_{ba} {\Aknow}^b_a  }
.
\label{eqn-cns-steady}
\end{align}
 Here we see that --
 as expected --
 if losses are small then each agent might achieve
 nearly perfectly accurate beliefs.
Conversely,
 if losses are large,
 then in this idealised steady state
 each agent is only left with the ``find by chance'' minimum ${\Akmin}$.
The threshold between these two extremes is 
 located in the regime where
 the effective information transmission rate ${\Trate}^b_a {\Elink}_{ba} {\Aknow}^b_a$ 
 becomes comparable to the information loss rate $\gamma {\Akmin}$.

\begin{figure}
\begin{center}
\resizebox{0.750\columnwidth}{!}{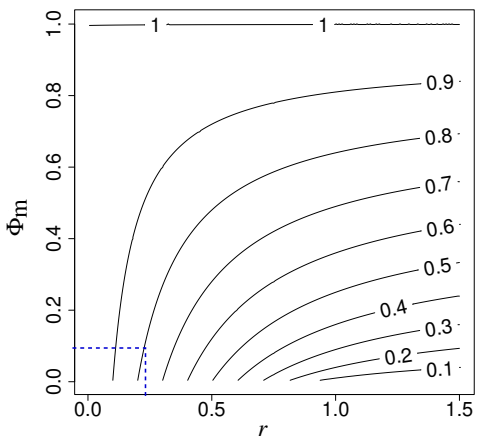}
\end{center}
\caption{{
Contour plot of the targeting accuracy ${\Aknow}^a_b$
 values from \eqref{eqn-cns-quadratic}, 
 as dependent on the find-by-chance probability ${\Aknow}_m$
 (which is usually set according to the scenario)
 and the effective accuracy loss rate $r=\gamma / {\Trate}^b_a {\Elink}_{ba}$. 
Note that the typical range of interest will be for small values
 of ${\Aknow}_m$,
 and small $r$;
 i.e. that at the lower left part of the figure.
As an example, 
 if $\Akmin=0.10$ then if we want to achieve a minimum 80\% accuracy
 then we need to keep the effective loss rate $r$ below $0.25$ 
 (as indicated by the dotted line); 
 for 90\% we would need $r<0.10$.
 If the information loss parameter $\gamma$ were fixed, 
  then to improve accuracies ${\Aknow}^a_b$
  we would need to increase the transmission rates ${\Trate}^a_b$.}
}
\label{fig-cns-quadratic}
\end{figure}

By using this $b \rightarrow a$ expression \eqref{eqn-cns-steady}
 in concert with its $a \rightarrow b$ counterpart, 
 we can get a quadratic expression for ${\Aknow}^a_b$
 that tells us the informational effect of any single agent-to-agent link
 as we show in the appendix.
If we then assume symmetric parameters,
 i.e. with 
 $r = \gamma / {\Trate}^b_a {\Elink}_{ba}  = \gamma / {\Trate}^a_b {\Elink}_{ab}$, 
 then we find that 
~
\begin{align}
  \left[
    {\Aknow}^a_b
  \right]^2
 +
  \left[
    r
   -
    1
  \right]
  {\Aknow}^a_b
 -
  r {\Akmin}
&=
  0
,
\label{eqn-cns-quadratic}
\end{align}
 which can be solved, 
 with the valid (i.e. positive valued) solution
 being plotted on fig. \ref{fig-cns-quadratic}.
We see that {in any scenario with
 a fixed minimum find chance $\Akmin$, 
 the performance
 (i.e. here the steady-state value of ${\Aknow}^a_b$)
 will degrade
 if the information environment becomes more challenging 
 (i.e. as loss $\gamma$ increases),}
 but with no sharp threshold behaviour.


%
\subsection{Communication tactics: payoffs and penalties}\label{S-geomfree-continuum-games}

If we make a simple assumption 
 that risk to an agent is simply proportional to 
 the message volume it sends, 
 any agent $a$ will wish to minimise its totalled $\sum_j{\Trate}^a_j$
 rates of data transmission.
Clearly, 
 therefore, 
 it might want to set ${\Trate}^a_b = 0$
 if it can reliably infer that ${\Elink}_{ab}$ is already small, 
 since this is an inefficient link and probably not 
 worth maintaining.
This optimization would be
 especially valuable if (e.g.) $a$ might instead communicate
 with $b$ via efficient links through some intermediate agent $c$.
However, 
 a low rate of incoming information ${\inforate}^a_b$
 could be due to any of three factors:
 (i) a low link efficiency ${\Elink}_{ba}$, 
 (ii) a sending agent with inaccurate beliefs ${\Aknow}^b_a$, 
 or
 (iii) a sending agent which has chosen a low transmission rate ${\Trate}^b_a$.

Of course,
 \emph{if} each agent $a$ has an estimate of ${\Akmin}$ and $\gamma$, 
 and is assumed to actually be aware of the values of 
 its set of ${\Aknow}^a_j$, 
 it could fix an $\alpha^a_j$,
 wait for steady state, 
 assume symmetric behaviour,
 and then attempt to estimate the ${\Elink}_{ja}$ for each $j$.
Given this, 
 it could customise its ${\Trate}^a_b$ accordingly, 
 although the assumption of symmetry --
 i.e. that the other agent(s) will be doing exactly the same thing,
 and at the same time --
 is a very aggressive one.

For example, 
 for an ER network \cite{Coscia-NetworkAtlas}
 in the limit where there is a large connected component,
 i.e. where the probability of an (${\Elink}=1$) link
 between any two agents is $p = \log(N)/N$, 
 a $1-p$ proportion of links might be pruned by such a process.
Then,
 assuming that all non-zero (and non-zeroed)
 transmission rates ${\Trate}^i_j$ have
 some constant value ${\Trate}_{\textup{c}}$, 
 we see that 
 the swarm risk rate likewise drops, 
 i.e. from proportional to ${\Trate}_{\textup{c}} N(N-1)^2$ to proportional to 
 ${\Trate}_{\textup{c}} N \log(N)$.
However,
 coordinating all the agents --
 and remember they may all have very different and possibly changing beliefs -- 
 so that they all prune the available links 
 down to compatible subsets 
 will be a tricky problem, 
 especially in this limited-communication regime.

%
\section{Agent and swarm metrics}\label{S-geomfree-metrics}

Before moving to the discrete communications model that 
 is used for the main results of this paper, 
 it is useful to consider some agent and swarm metrics
 that can be used to judge how
 an agent or swarm instance is performing.

Note that in the following we assume that the inter-agent messaging
 does not include agents forwarding any information 
 about others; 
 i.e. a message sent from $a$ to $b$ will not also 
 contain information about a third agent $c$.
This is a restrictive assumption, 
 but one which greatly simplifies the model
 (cf. with \cite{Kinsler-EHMW-2022anglen}).
It enables us to set some benchmarks for agent behaviour 
 without introducing the considerable complications 
 of how an agent might manage -- 
 and make inferences from --
 {a diverse array of partial,
 uncertain, 
 and variously out-of-date information
 about (e.g.) the locations of the other agents.}

%
\subsection{Messaging rates}\label{S-geomfree-metrics-norm}

It is useful to assume that our agents have some maximum 
 total transmission rate $A$ that they are capable of; 
 this also helps us set timescales on 
 which the dynamics occurs, 
 as well as assist conversion into the discrete model
 treated later.
Thus we want to normalise ${\Trate}^a_j$
 so that the \emph{total} information transmission rate
 conforms to 
~
\begin{align}
 \Trmax
&\ge 
 {\Trsum}^a
=
 \sum_j {\Trate}^a_j
.
\end{align}
We see here that 
 an agent is not required to transmit at the maximum rate $\Trmax$, 
 and it can instead transmit at some reduced rate ${\Trsum}^a$.

If one imagined that an agent $a$ could infer a value
 for ${\Elink}_{ab}$, 
 {it could
 set ${\Trate}^a_b$ to zero if it believed ${\Elink}_{ab}$ too small
 to be worth trying to overcome; 
 or increase ${\Trate}^a_b$ if it believed ${\Elink}_{ab}$ large enough}
 to support a useful information flow.

%
\subsection{Performance}\label{S-geomfree-metrics-performance}

A simple measure of how well informed an agent $a$ is 
 might be 
 constructed 
 by simply summing some ``performance'' function of the belief accuracies ${\Aknow}^a_j$
 and then applying some appropriate normalisation.
The idea here is that the closer the performance measures suggested below
 are to unity, 
 the more likely it is that the agents have sufficient good information
 with which to communicate effectively.

A simple link performance function of ${\Aknow}^a_j$ might be $({\Aknow}^a_j)^\beta$, 
 perhaps with just $\beta=1$,
 but where choosing $\beta>1$ will de-emphasise inaccurate beliefs in the measure.
In this case, 
 where the performance measure includes contributions from 
 all ${\Aknow}^a_j$ values, 
 the normalisation should simply be $1/N$.
Thus the agent performance measure is
~
\begin{align}
  \bar{\Aknow}^a(\beta)
&=
  \frac{1}{N}
  \sum_j
    \left( {\Aknow}^a_j \right)^\beta
,
\end{align}
 and by extension the swarm performance measure is
~
\begin{align}
  \hat{\Aknow}(\beta)
&=
  \frac{1}{N}
  \sum_i
    \bar{\Aknow}^i
\quad
=
  \frac{1}{N^2}
  \sum_i
  \sum_j
    \left( {\Aknow}^i_j \right)^\beta
.
\end{align}

However, 
 this doesn't work well for scenarios where
 each agent $a$
 doesn't necessarily need to be aware of \emph{every} other agent $j$,
 just a subset with hopefully good link efficiency $L_{aj}$,
 that enables reliable connectivity over a sufficiently small number of hops.
In such a case we might choose a suitable threshold value $\pthreshold$
 for the accuracy, 
 and only include the ``good'' contributions, 
 i.e. those where ${\Aknow}^i_j > \pthreshold$.
That is, 
 using the Heaviside step function $H(x)$, 
 we set the performance measure for an agent $a$ to be 
 based on
~
\begin{align}
  h^a_j( {\Aknow}^a_j ; \pthreshold)
&=
  {\Aknow}^a_j
  ~
  H( {\Aknow}^a_j  -   \pthreshold )
.
\end{align}
 and the normalisation $\bar{N}^a$ as
 the \emph{maximum} of two possible values:

\begin{description}

\item[(i)]
 a sum over accurate links,
 i.e.
~
\begin{align}
  \bar{N}^a ({\Aknow}_{\upT})
&=
  \sum_j
    H( {\Aknow}^a_j  -   \pthreshold )
,
\end{align}
  or

\item[(ii)]
the 
 ER average number of links per node
 when in the large connected component limit, 
 i.e.
~
\begin{align}
 \tilde{N}^a = \log(N).
\end{align}

\end{description}

\begin{figure}
\begin{center}
\resizebox{0.750\columnwidth}{!}{\includegraphics{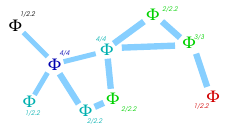}}
\end{center}
\caption{{
Example calculation of the $\hat{\Aknow}_{\upT}$ performance metric.
For this 9-agent swarm, 
 with accurate links indicated with thick light blue lines, 
 the ``just connected'' ER expected number of links per agent 
 is $\log(9) \simeq 2.2$; 
 thus when working out the link-count normalisation 
 we use this value even if the actual number
 of accurate links is 0, 1, or 2.
Summing the contribution from each agent 
 we get $(1+1+2+2+2+1)/2.2~+~(1+1+1) = 7.1$, 
 for a swarm performance measure of $\hat{\Aknow}_{\upT} \simeq 7.1/9=0.79$.
}}
\label{fig-performancy}
\end{figure}

This ``maximum of'' is used as a pragmatic way 
 to counteract misleading cases where 
 an agent has (or agents have) too few beliefs that are sufficiently accurate
 to suggest any swarm connectivity, 
 but where those that are accurate are nevertheless well above threshold,
 and so would otherwise return a misleadingly high performance measure.

Thus this thresholded agent performance measure
 $\bar{\Aknow}^a_{\upT}$ is
~
\begin{align}
  \bar{\Aknow}^a_{\upT}
&=
  \frac{1}{\bar{N}^a}
  \sum_j
    {\Aknow}^a_j
  ~
    H( {\Aknow}^a_j  -   {\Aknow}_{\upT} )
,
\end{align}
 and by extension the corresponding 
 swarm performance measure is 
~
\begin{align}
  \hat{\Aknow}_{\upT}
&=
  \frac{1}{N}
  \sum_i
    \bar{\Aknow}^i
\quad
=
  \frac{1}{N}
  \sum_i
    \frac{1}{\bar{N}^i}
  \sum_j
    {\Aknow}^a_j
    H( {\Aknow}^a_j  -   {\pthreshold} )
.
\end{align}
{An example calculation is indicated on fig. \ref{fig-performancy}.}

Note that
 for any performance measure, 
 whilst an agent $a$ might calculate (or estimate)
 its own performance $\bar{\Aknow}^a$, 
 it cannot calculate the swarm performance $\hat{\Aknow}$.

As an aside,
 for any single set of link efficiencies $\{ {\Elink}_{ij} \}$,
 {we could compute a customised performance scheme 
 that uses information about what actual links are good,
 as opposed to the approaches introduced above where we attempted a 
 reasonable and \emph{general} normalisation.
However, 
 since an agent is never aware of the true values of  
 $\{ {\Elink}_{ij} \}$,}
 such a custom performance measure is not calculable by an agent attempting 
 to determine whether it has sufficient good information 
 about a sufficient number of nearby others.
Nevertheless, 
 such an omniscient measure could still be used as a benchmark
 for comparing other agent performance measures.

%
\subsection{Connectedness}\label{S-geomfree-metrics-connected}

\begin{figure}
\begin{center}
\resizebox{0.750\columnwidth}{!}{\includegraphics{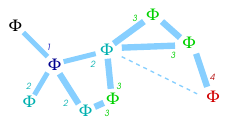}}
\end{center}
\caption{{
Counting the number of hops needed to pass a message 
 from one agent to another.
Here accurate links are indicated by the thick light blue lines, 
 and these links are also assumed to be able to pass messages efficiently.
In the situation shown here, 
 although most agents are easily within three hops of one another, 
 the two on the left and right extremes
 are four hops away from each other.
Thus this swarm does not satisfy the CCS criteria,  
 but it would if the dashed link were \emph{also} accurately known.
However, 
 since it is the cumulative probability that counts for connectivity, 
 many hop paths taken over imperfectly accurate links will 
 in combination be less accurate
 (being the product of each hop accuracy), 
 and may fall below the CCS criteria for being connected.  
}}
\label{fig-connectify}
\end{figure}

A ``completely connected swarm'' (CCS)
 is formed if it is possible to send a message 
 from any one agent to any other agent, 
 possibly via intermediate agents.
Here, 
 this determination is also subject to there being 
 (a) no more than three such hops, 
 and 
 (b) that the probability of the message successfully traversing
 the whole path is above some suitably chosen probability threshold  $\pthreshold$; 
 {as indicated on fig. \ref{fig-connectify}.}
Here we typically choose $\pthreshold = 0.75$
 and three hops, 
 {as convenient criteria
 that provide representative results,
 and making CCS connectivity achievable, 
 but not always guaranteed.}
This is a different measure than the 
 large connected component
 of standard network theory, 
 but is chosen to mimic plausible limitations
 on message passing across the swarm.
{That is, 
 this choice of $\pthreshold= 0.75$ is chosen
 so as to cover scenarios where
 some message loss is present,
 but where messages are not routinely lost.
Increasing the value of $\pthreshold$ demands
 more reliable messaging, 
 and makes it harder to achieve the CCS criteria; 
 decreasing it makes it easier; 
 but in general the results are not particularly sensitive
 to any specific choice.}

This determination of connectedness does make
 the artificial assumption that the complete set of accuracies ${\Aknow}^i_j$
 could somehow be collected together, 
 to enable a matrix of agent-to-agent message success probabilities.  
Thus, 
 whilst a useful metric for judging the success
 of a simulation or some chosen messaging tactics, 
 it is not something that any one agent can calculate.

Indeed,  
 in this simple model here we have that each agent $a$
 is only aware of its own star-network; 
 it has no way of inferring anything 
 about connected components that involve 
 hopping along multiple links.
Any non-star ``connected component'' is 
 not something an agent can ever be aware of,
 and therefore is not something an agent
 can use to make decisions.
This is why we introduced the agent-centric
 performance measures $\bar{\Aknow}^i$
 above, 
 {since these tell us whether each
 agent is likely to have enough accurately linked neighbours}
 in its star, 
 so that a CCS 
 (or even a large connected component)
 might exist.

Further, 
 we \emph{do not} include the effect of the link efficiencies ${\Elink}_{ij}$, 
 because here agents are always unaware of their values,
 and they are not estimated either.
Clearly, 
 this lack might be problematic, 
 but if the ${\Elink}_{ij}$ all have values either near 1 or 0, 
 the ${\Aknow}^i_j$ should also tend to these values, 
 so the discrepancy should be manageable.

%
\subsection{Risk}\label{S-geomfree-metrics-risk}

The rate $\bar{\Rrisk}^a$ that risk accumulates to any agent $a$ in this model 
 is most simply assumed to be in proportion
 to a sum over its transmission rates ${\Trate}^a_j$.
I.e., 
 we have
~
\begin{align}
  \bar{\Rrisk}^a
&=
  \xi
  \sum_j
    {\Trate}^a_j
,
\end{align}
{where $\xi$ is a proportionality constant
 that converts a messaging rate into 
 the rate of detection by the adversary.}
As a result,
 we can clearly see that the maximum transmission rate $\Trmax$
 set previously then also sets a maximum risk accumulation rate (``risk rate'') 
 of $\xi \Trmax$.
However, 
 an agent is not required to transmit at this maximum rate, 
 so the actual risk rate can be smaller.

{For all agents (i.e. the swarm),
 we then have the total}
~
\begin{align}
  \hat{\Rrisk}
&=
  \xi
  \sum_i
  \bar{\Rrisk}^i
\qquad
=
  \xi
  \sum_i
  \sum_j
    {\Trate}^i_j
.
\end{align}

Note that for any risk measure, 
 whilst an agent $a$ might calculate its own risk rate $\bar{\Rrisk}^a$, 
 it cannot calculate the swarm risk rate $\hat{\Rrisk}$.
Further, 
 these risk rates are not the same as detection probabilities
 except in the limit where they are small, 
 i.e. when $\ll 1$.
Instead, 
 for some risk rate $R'$ over an interval $\tau$
 {we can calculate the detection probability to be}
~
\begin{align}
  p 
&=
  1 
 - 
  \exp \left( - R' \tau \right)
.
\end{align}

%
\section{Discrete communications}\label{S-geomfree-stochastic}

Although many of the parameters in this model are identical 
 to the continuum model above, 
 our discrete communication model 
 has stochastic transmission and reception algorithms. 
This communications model is one which more closely matches 
 a realistic situation of discrete messages 
 being sent one after another to 
 variously selected target agents.
{To simulate this system we wrote and used a bespoke fortran 
 code designed around this model and its planned extensions.}

Here we assume that time passes in discrete steps
 (or ``ticks'' of duration $\tau$ each), 
 and that in each tick each agent 
 could at most transmit one message 
 to any one other agent.
For example, 
 a trivial comunications tactic might be to select the 
 target agent $b$ at random from the list of other agents.
{This discrete communications model 
 is very different to the continuum model which sends a
 continuous trickle}
 of information to multiple other agents simultaneously.

\def\Trgoal{\Trsum^*}

Simulation parameters are set so that each agent 
 sends a maximum of one message per tick, 
 {but over any period of $\nticks$ ticks,
 we set the goal of sending
 $N$ messages.
This goal of sending at a rate $\Trgoal = N /\tau \nticks$
 is the counterpart of the ``$\Trmax$'' parameter from the continuum model.
Note that we need $\nticks > N$ to ensure that the maximum allowed
 messaging rate can be maintained, 
 although a greater margin is preferable 
 so that we can neglect the possibilty of congestion effects.
The message rate is reduced from its 1-per-tick maximum by applying conditions}
 which need to be met before any message is sent, 
 as we describe later in Sec. \ref{S-geomfree-stochastic-tactics}.

For reception, 
 the link efficiency ${\Elink}_{ab}$ 
 and the sender's available information ${\Aknow}^a_b$
 can be used as a probabilistic filter
 to determine whether a message was received.
If luck is on the receiver's side,
 the message arrives successfully and
 it gets all the information; 
 if not, 
 then it gets none.

%
\subsection{Updates (dynamics)}\label{S-geomfree-stochastic-updates}

The resulting update equations
 for a tick of length $\tau$ 
 can now be separated into two stages.
For the loss update
 we alter each agent $a$'s information vector ${\Aknow}^a_j$
 according to
~
\begin{align}
  {\Aknow}^a_j
&\leftarrow
 {\Akmin}
 +
  \exp\left( - \gamma \tau \right)  
  \left( {\Aknow}^a_j - {\Akmin} \right)
,
\end{align}
 and for the communications stage 
 we update according to
~
\begin{align}
  {\Aknow}^a_j
&\leftarrow
  {\Aknow}^a_j
 +
  W^j_a
  .
  \left( 1 - {\Aknow}^a_j \right)
,
\end{align}
 {where in any one tick, 
 the $W^j_a$ is essentially a yes-or-no 
 (i.e. 1 or 0)
 list of all 
 successful receptions at $a$ of a (possible) transmission
 from all $j$.
It is useful to split this $W^j_a$ into two parts,
 with $W^j_a = w^j_a b^j_a$, 
 where $w^j_a$ and $b^j_a$ are defined next.}

The first part ($w^j_a$) contains a ``1'' only if 
 {any agent $j$ transmissions sends a message to target $a$.}
Note that 
 for any sending agent $b$, 
 only the entry for
 the message's intended target $a$
 in $w^b_i$ is non-zero (i.e. $w^b_i=0$, except for $i=a$, since $w^b_a=1$).
\emph{Which} agent is chosen as a target in any given tick 
 is now dependent on the chosen communications tactic
 (see below in Sec. \ref{S-geomfree-stochastic-tactics}), 
 and the (average) probability with which an agent $a$ is chosen as a target by $j$
 is -- for small probabilities --
 related to the quantity $\tau {\Trate}^j_a$ from the continuum model.

The second part ($b^j_a$) is 
 the reception filter, 
 which will only be non-zero (i.e. unity)
 if the transmission is successfully received.
 i.e.
 if some random number $\eta^j_a$ chosen from a uniform distribution on $[0,1]$
 is such that $\eta^j_a < {\Elink}_{ja} {\Aknow}^j_a$.

This means that in any given tick,
 $W^j_i = w^j_i b^j_i$ will be mostly zeroes, 
 and have \emph{at most} $N$ entries that are one, 
 and then only all $N$ if each sent transmission
 is lucky enough to successfully pass the reception filter.
So here we see that unlike the continuum model's 
 gradual handling of information
 arrival and accumulation, 
 in each tick here only some information elements are updated, 
 and if they are updated, 
 they are updated to become perfectly accurate beliefs,
 i.e. if $a$ receives from $b$,
 then it sets ${\Aknow}^a_b=1$.

Regarding the normalisation, 
 performance,
 and risk considerations present for the continuum model, 
 here we have that:

\begin{enumerate}

\item
{The normalisation needed in the discrete model is more complicated
 than in the continuum model, 
 where we could just set a maximum rate $\Trmax$,
 i.e. just $\Trmax\tau$ messages per tick.
In the discrete model there is an absolute maximum 
 for each agent of
 at most one message per tick, 
 but as described above we usually want to send fewer than this, 
 and so set a goal of (at most) $\Trgoal$, 
 i.e. $N$ messages per time $\tau \nticks$.
Thus the relevant comparison is between the continuum model's $\Trmax$
 and the discrete model's goal of $\Trgoal$.}

\item
The performance measure(s) defined for our continuum model
 can be reused here,
 since they only depend on ${\Aknow}^i_j$.
However, 
 since we now need to run large ensembles of instances of this model,
 we are not restricted to only considering only one outcome, 
 or an averaged behaviour,
 of the ${\Aknow}^i_j$ and values derived from it.
We can also consider their distributions
 as accumulated over both time intervals 
 and the many different instances.

\item
The risk measure(s) defined for our continuum model
 can be reused here,
 but here they reduce to a simple message counting.
To go beyond this requires both a signalling model 
 and an adversary model
 to be specified (e.g. see \cite{CEME-DASA-2022}).

\end{enumerate}

%
\subsection{Communication tactics}\label{S-geomfree-stochastic-tactics}

In this discrete communications approach, 
 we can no longer message all other agents at once, 
 but can only send a message to one target at a time. 
This situation immediately suggests the simple tactic 
 of targetting each agent sequentially, 
 i.e. one-by-one in some fixed order; 
 but other simple schemes are possible,
 most notably just choosing targets at random.
More complicated tactics could involve 
 choosing targets based on what messages were received
 and their timings.
{Any agent $a$, 
 however, 
 will need to consider the balance between 
 the average number of messsages sent to other agents $i$ per tick 
 (the counterpart of $\tau {\Trate}^a_i$ from the continuum model)
 and the belief decay rate ($\gamma$), 
 and its effect on the performance of the other agents.}

In the continuum case, 
 each agent is aware of its transmission rates ${\Trate}^a_i$
 and the receive rates ${\inforate}^a_j$.
{Here, 
 in analogy,
 agent $a$ is aware of 
 (i) $\mathscr{T}^a_j$, 
      how many ticks ago they last sent a message to the agent $j$, 
 and
 (ii) $\mathscr{R}^a_j$, 
     how many ticks ago they last received a message from the agent $j$.}
Further, 
 each transmitted message contains the sender's matching receive-from time, 
 so that a messaging there-and-back ``round-trip''
 $\mathscr{O}^a_j$ time can be calculated.

Although there is a large range of 
 possible communications tactics, 
 here we will restrict ourselves to a short list,
 three of which are closely related.
{We aim to address more sophisticated tactics
 in future work.}
The communications tactics we consider ensure that 
 the messages of a sending agent 
 are directed 

\begin{enumerate}[label=(\roman*)] 

\item \label{enum-tactic-circ}
 towards each other agent in a fixed and repeating \textit{Sequence};

\item \label{enum-tactic-rand}
 towards another agent chosen at \emph{Random};

\item \label{enum-tactic-time}
 towards another agent chosen at random, 
 as long as a \emph{Timer} threshold has been exceeded;

\item \label{enum-tactic-bounce}
 towards another agent chosen at random, 
 as long as a {Timer} threshold has been exceeded; 
 except if that target agent has not reported 
 sufficiently recent contact from the sender
 (\emph{Filtered});

\item \label{enum-tactic-bounceplus}
 as per {Filtered}, 
 but with an extra probability of a small
 number sent towards another agent chosen at random
 (\emph{Filtered+}, \emph{Filtered++}).
Filtered+ incorporates a 25\% chance of considering such an extra message,
 whilst Filtered++ has 50\%.

\end{enumerate}

The first three tactics here --
 i.e.
 \ref{enum-tactic-circ}, \ref{enum-tactic-rand}, \ref{enum-tactic-time} --
 do not attempt to select or deselect 
 targets except as a rate-management technique; 
 even inefficient or unreliable links will continue to be used.
Although this does make them robust to 
 unexpected changes in link efficiency, 
 it also means they will waste messages
 and increase risk unnecessarily 
 by transmitting over inefficient links.
Nevertheless, 
 they provide a set of baseline cases
 which we should be able to outperform.

The Filtered tactic \ref{enum-tactic-bounce}
 proposed here adapts to the environment by 
 excluding poor links. 
Its key feature is that it 
 does this \emph{without} requiring inferences 
 about ${\Aknow}^a_i$ or ${\Elink}_{ai}$ values.
Only timing data is needed,
 {and for a proposed message from sending agent $a$ to target agent $b$, 
  the only non-trivial requirement is that 
  the last message from $b$ received by $a$
  included the time $\mathscr{R}^b_a$.}
This tactic, 
 and its derivatives, 
 are the only ones considered here that will on average send fewer messages
 {than the goal rate $\Trgoal$,}
 as excluded messaging possibilities are not 
 redirected elsewhere, 
 but dropped.
However, 
 once links have been deemed bad and dropped, 
 they will never be recovered by this Filtered tactic.
This is why we also introduce the 
 Filtered+ and Filtered++ tactics 
 \ref{enum-tactic-bounceplus}
 which --
 as we will see --
 delay this process.

%
\section{Results: static environment}\label{S-estatic}

We now evaluate the performance of our chosen communications tactics
 using the discrete communications approach.
We do not model the continuum version, 
 since its ``continually transmit to everyone, always''
 approach is unlikely to be of much practical use; 
 even if it did serve a useful function in presenting 
 some basic concepts.

Here we will consider a swarm of $N=20$ agents
 and contrast the cases of an in-principle fully-connectable swarm
 (i.e. where all ${\Elink}_{ij} \sim 1$)
 and a partially-connectable swarm
 where ${\Elink}_{ij}$ is a mix of zeroes and ones.
The  ``hit by chance'' minimum message targetting probability ${\Akmin}=0.10$, 
 which, 
 if imagining a set of spatially distributed agents,
 is compatible with a \emph{directed} transmissions
 covering an angle of $36^{\circ}$.
However, 
 in this model 
 these messages are only ever received by their intended recipient, 
 or they are not received at all.
Messages are never received accidentally by some other (non-targetted) agent, 
 which might happen in a true spatial description
 \cite{CEME-DASA-2022,Kinsler-EHMW-2022anglen}.

Each simulation run starts 
 with randomly chosen transmitted-to 
 and received-from communications timings, 
 thus setting up a plausible initial state.
Nevertheless, 
 since this is not guaranteed to be a good match to 
 a typical quasi-steady state, 
 we start by propagating for over 10 thousand (10k) ticks so that the distribution
 of agent properties should no longer be dependent on the initial conditions, 
 and the evolution can be reasonably assumed to be ergodic.
Then the next $\sim 20$k ticks are averaged over
 in lieu of averaging over multiple independent simulations.

%
\subsection{Fully connectable swarm}

In a fully connectable swarm, 
 every agent has a good chance of transmitting successfully to 
 each other agent, 
 i.e. ${\Elink}_{ij} \sim 1$.
This makes it easy to successfully achieve a large,
 all-agent connected component, 
 but there will likewise be a large number
 of good -- 
 but unnecessary -- 
 links that an agent might think worth maintaining.

Specifically, 
 we consider a ``flat'' environment
 where ${\Elink}_{ij} = 0.95$, 
 i.e. that there is only a small efficiency loss, 
 and there are no preferred (or excluded) agent-to-agent links.
Since all links are (initially) allowed for use 
 by any of the communications tactics, 
 and links are equally efficient, 
 all-averaged measures --
 e.g. averaged ${\Aknow}^a_b$ or averaged functions thereof
 are useful as a test of performance.

\begin{figure}
\resizebox{0.4800\columnwidth}{!}{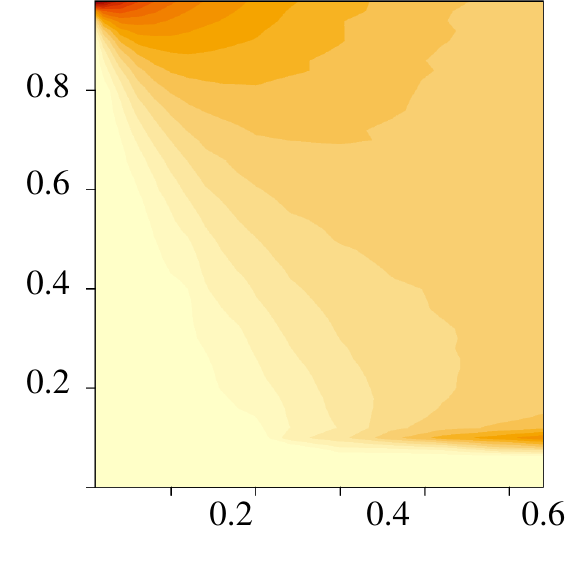}
\resizebox{0.4800\columnwidth}{!}{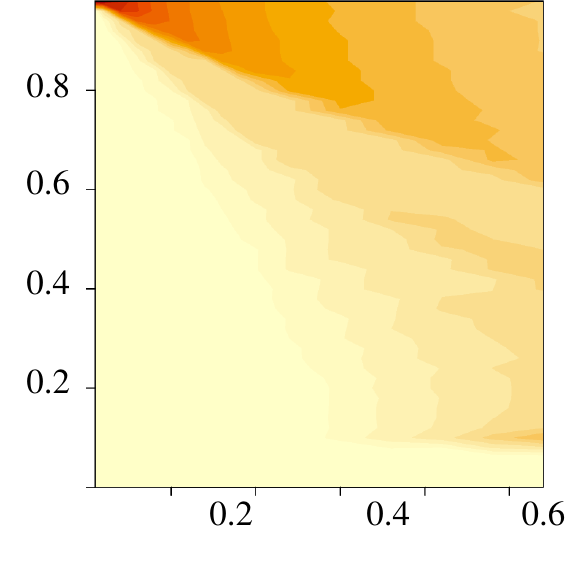}
\caption{
Filled contour plot based on histograms of the occurence-frequency 
 of binned values of ${\Aknow}^i_j$
 and as a function of $\bar{\gamma}=\gamma \tau \nticks$.
{Other parameters are 
 $N=20$, $K = 200$, 
 ${\Akmin}=0.10$,
 with all ${\Elink}_{ij} = 0.95$.}
Each simulation, 
 for {the resulting} agent-independent ${\Trate}^i_j = \alpha$, 
 produces one histogram that makes up
 one vertical slice of the whole figure panel; 
 thus we can scan the figure left-to-right and see how
 the behaviour will change as this loss ratio is increased.
Here the color-coding,
 where dark stands for more likely,
 is based on 
 the \emph{cube root} of the occurence-frequencies
 so as to improve feature visibility.
On the left,
 we see the results using the Random tactic \ref{enum-tactic-rand};
 and on the right,
 the Timer tactic \ref{enum-tactic-time}.
The results for the Sequential tactic \ref{enum-tactic-circ}
 are very similar to those for Timer \ref{enum-tactic-time}.
}
\label{fig-results-flat-simple-hh}
\end{figure}

Fig. \ref{fig-results-flat-simple-hh}
 shows a transition from a ``good swarm''
 with much high-probability and accurate agent beliefs,
 and, 
 as the normalised decay rate increases, 
 changing to a fragmented (non) swarm 
 with only the floor probability ${\Akmin}$ remaining.
There is a transition region
 where the swarm starts to fail,
 where the accuracy values of ${\Aknow}^a_b$ are more widely distributed.
Here we see that the Timer tactic performs better
 than the Random one, 
 as the frequency of high-value ${\Aknow}^i_j$ is better.

\begin{figure}
\begin{center}
\includegraphics[width=0.750\columnwidth]{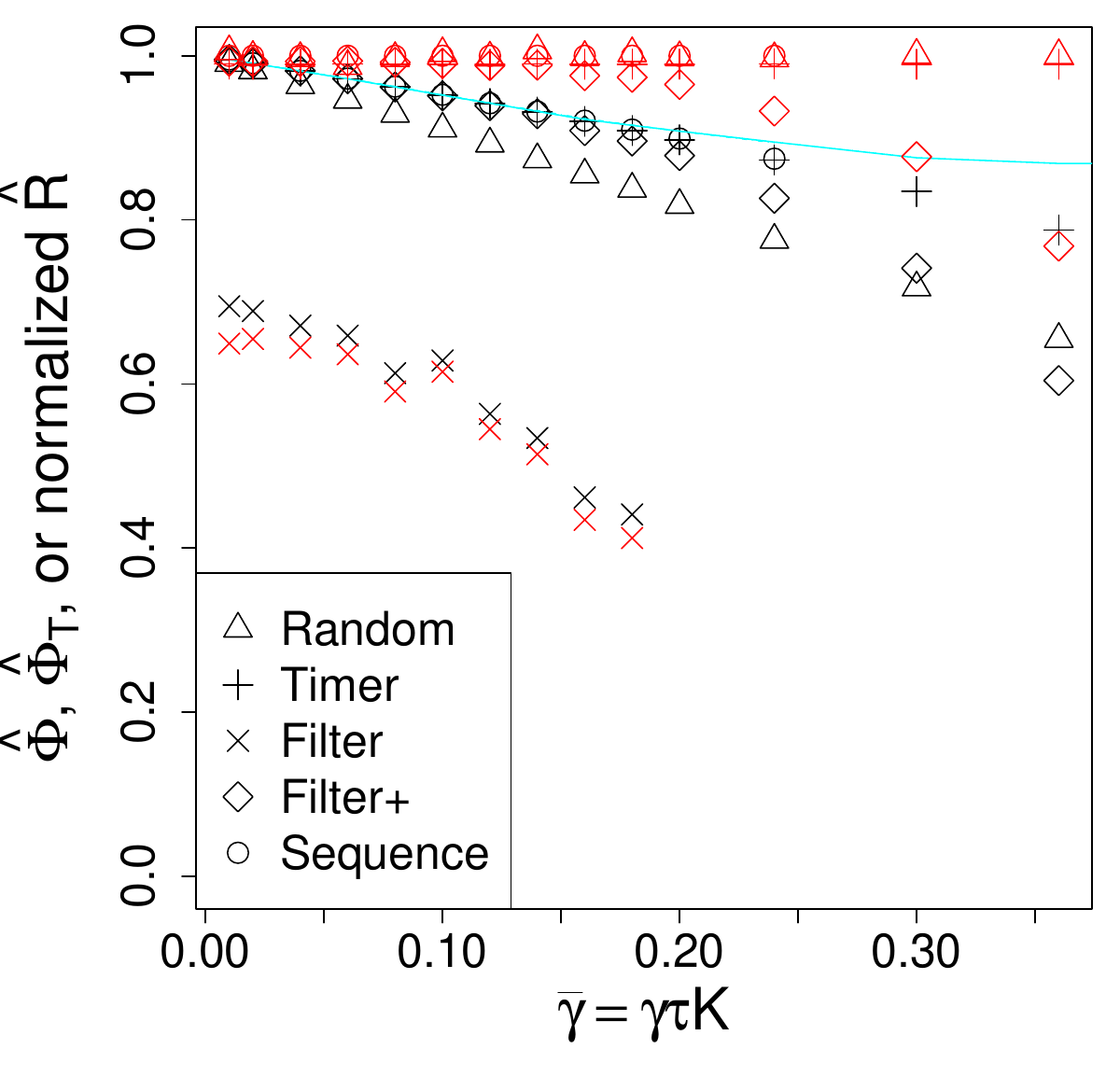}
\end{center}
\caption{
Comparisons of performance
 in both $\hat{\Aknow}(1)$ (black symbols)
 and normalised risk $\hat{R}$ (red symbols)
 for the four (five) communications tactics.
The ``missing'' $\times, \circ$ points at high $\bar{\gamma}$ 
 are omitted for cases where no CCS was formed.
The cyan line is that for the $\hat{\Aknow}_{\upT}$ measure, 
 which in this plot is nearly the same for every tactic.
Note that $\bar{\gamma}$ is the counterpart of $r$
 in the continuum model.
}
\label{fig-results-flat-cfar}
\end{figure}

In fig. \ref{fig-results-flat-cfar}
 we see that in terms of swarm performance, 
 the Random communications tactic is typically the  worst.
We should expect this, 
 since some links will,
 by chance,
 and albeit temporarily,
 be neglected by an agent, 
 leading to its accuracy dropping further.
In contrast, 
 the Timer and Sequence tactics are more regular
 in messaging over links, 
 leading to a more compact distribution of accuracies
 (see fig. \ref{fig-results-flat-simple-hh}), 
 and higher values
 (fig. \ref{fig-results-flat-cfar}).
However, 
 all three tactics 
 have a similar risk profile because they
 are subject to the same rate limiting, 
 and do not ignore any link posssibilities.

The results for the Filtered tactic here are
 only indicative, 
 and not strictly in a quasi-steady state.
This is 
 because this tactic chooses to prohibit sending on some links, 
 but will never re-enable them, 
 so  that even good links will --
 in principle -- 
 eventually have a sufficiently unlucky period
 that results in them being switched off.
An all-links switch-off 
 can be seen at about 30k ticks into the simulation, 
 but only 
 at extremely high losses $\bar{\gamma} >0.45$, 
 is still in-progress for $0.23 < \bar{\gamma} < 0.45$, 
 but not yet evident for $\bar{\gamma} < 0.23$.
The Filtered+ and Filtered++ tactics, 
 with their additional random message targets, 
 significantly delay this process,
 but do not entirely stop it.


%
\subsection{Partially connectable swarm}

\begin{figure}
\begin{center}
\includegraphics[width=0.750\columnwidth]{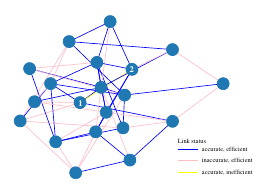}
\end{center}
\caption{
{Background environmental network as defined by ${\Elink}_{ij}$, 
 shown with link colours indicating the final simulation state
 of a simulation with a relatively large loss $\bar{\gamma} = 0.24$,
 and using the Filter+ tactic. 
Blue lines are links that are both efficient (environmental)
 and that are known accurately,
 whereas the light pink ones 
 are efficient but inaccurately known 
 (i.e. ``lost'' links''); 
 the partly obscured yellow line is an 
 inefficient link between agents $1$ and $2$
 that by luck has unexpectedly become accurate.
Although each agent in this diagram 
 is indeed linked to each other agent 
 via intermediate agents,
 for some of those paths \emph{more} than three hops
 are required, 
 and so a CCS is not present here.}
}
\label{fig-results-back-cfnet}
\end{figure}

Next we consider a partially connectable swarm, 
 i.e. we choose an 
 environment where ${\Elink}_{ij}$
 is randomly chosen to be either zero or one, 
 forming a swarm of agents that are connected
 to only a few neigbours, 
 but which are connected enough so that 
 they can in principle still be globally connected.
Here we generate the same random network
 for each different loss value tested, 
 with a link probability of $p=0.284$, 
 {and one instance of the resulting connectivity}
 can be seen in fig. \ref{fig-results-back-cfnet}.
The idea here was to ensure that the swarm was 
 better connected than a ``just connected'' ER network
 with $p=\log(N)/N$, 
 {which for $N=20$ gives $p\simeq 0.150$,}
 so that there are some redundant links, 
 but not too many.
{This boost in link probability
 also helps improve connectivity,
 since at $N=20$ we are not in the 
 large $N$ limit where the ER criteria is valid.}

\begin{figure}
\resizebox{0.4800\columnwidth}{!}{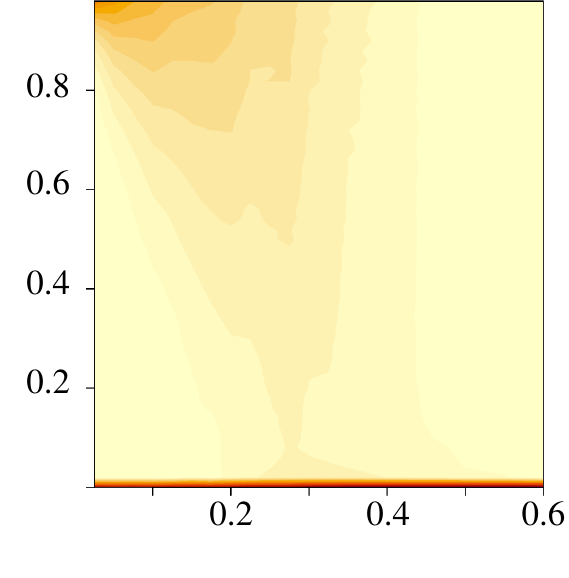}
\resizebox{0.4800\columnwidth}{!}{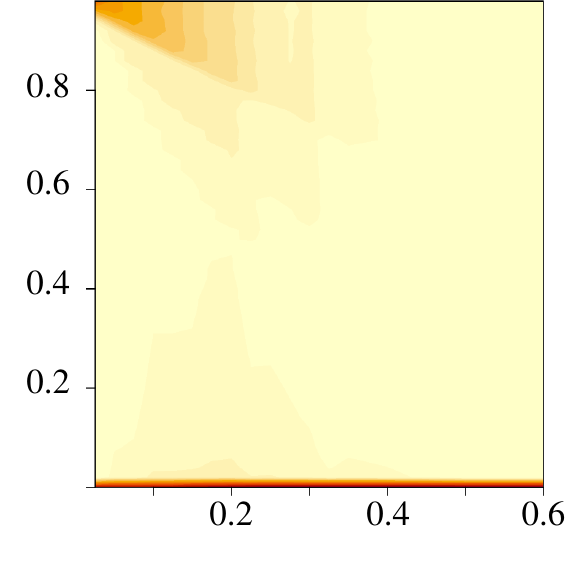}
\caption{
Filled contour plot based on histograms of the occurence-frequency 
 of binned values of ${\Aknow}^i_j$
 and as a function of rate ratio $\bar{\gamma}=\gamma \tau \nticks$.
{Other parameters are 
 $N=20$, $K = 200$, 
 ${\Akmin}=0.00$,
 with the ${\Elink}_{ij}$ being randomly 0 or 1 
 as described in the main text.}
Each simulation, 
 for some $\bar{\gamma}$,
 produces one histogram that makes up
 one vertical slice of the whole figure panel; 
 thus we can scan the figure left-to-right and see how
 the behaviour will change as loss is increased.
Here the color-coding,
 where dark stands for more likely,
 is based on 
 the \emph{cube root} of the occurence-frequencies
 so as to improve feature visibility.
On the left,
 we see results using the Random tactic \ref{enum-tactic-rand},
 and on the right,
 the Filtered tactic \ref{enum-tactic-time}. 
The results for Timer and Sequential tactics
 are very similar to Filtered, 
 but with better high $\bar{\gamma}$ performance.
}
\label{fig-results-back-simple-hh}
\end{figure}

Fig. \ref{fig-results-back-simple-hh}
 show transitions away from a ``good swarm''
 with much high-probability and accurate agent beliefs,
 through to a fragmented (non) swarm 
 with only the floor probability ${\Akmin}$ remaining.
Compared to fig. \ref{fig-results-flat-simple-hh},
 the good swarm regime persists for a smaller range
 of information decay strengths, 
 as might be expected given the much reduced number
 of efficient links.

\begin{figure}
\begin{center}
\includegraphics[width=0.70\columnwidth]{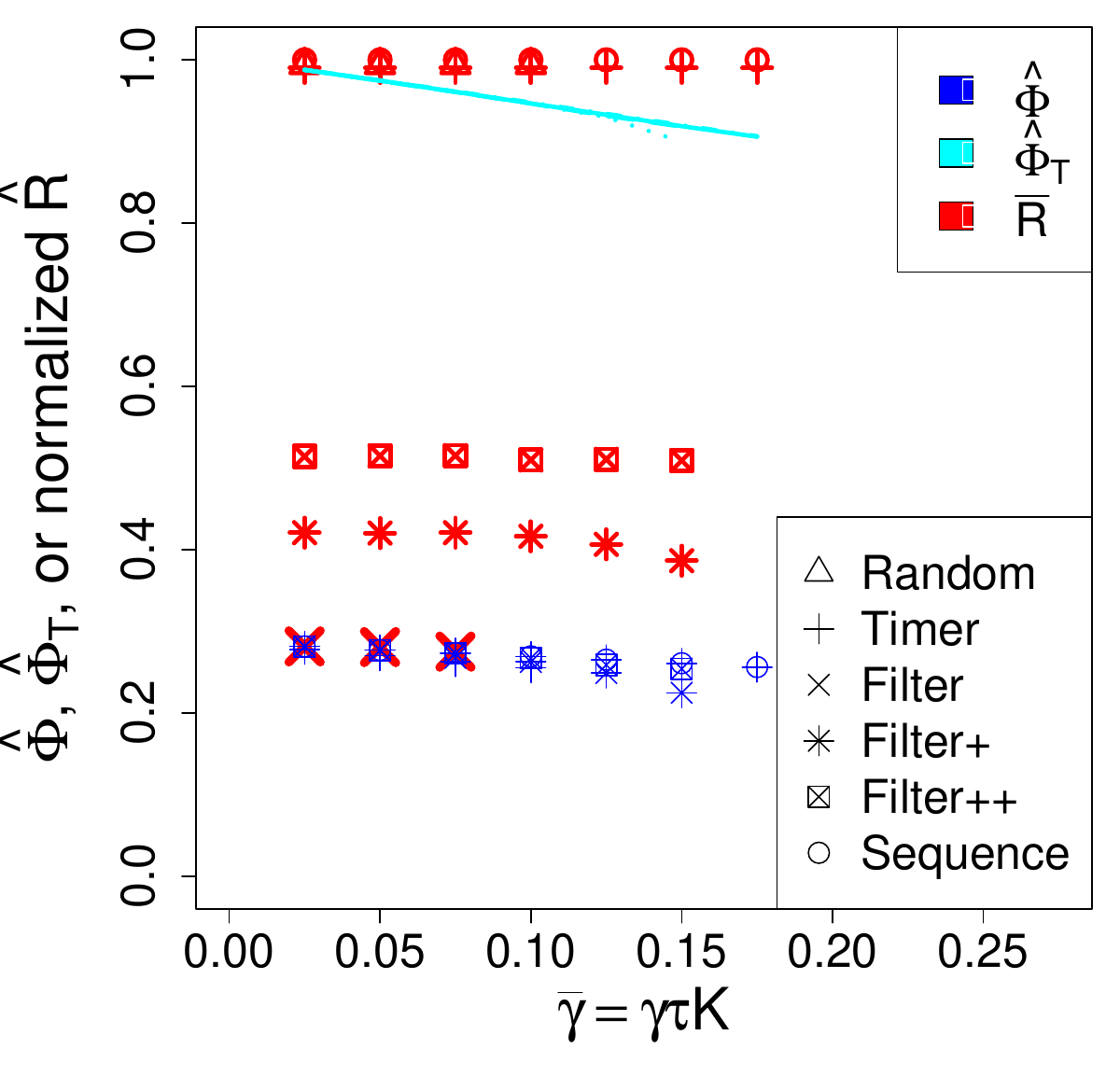}
\end{center}
\caption{
Comparisons of performance
 in both $\hat{\Aknow}(1)$ (blue symbols)
 and scaled risk $\hat{R}$ (red symbols)
 for the four (five) communications tactics.
The ``missing'' points at higher $\bar{\gamma}=\gamma \tau \nticks$ 
 are omitted for cases where no CCS was formed.
The cyan line(s) are those for the $\hat{\Aknow}_{\upT}$ measure, 
 which in this case are nearly the same for every tactic, 
 and here are \emph{the} most relevant metric for swarm performance.
}
\label{fig-results-back-cfar}
\end{figure}

Since only a fraction of the ${\Elink}_{ij}$
 are non-zero, 
 the simple performance measure ${\hat{\Aknow}}$
 is low, 
 because it reflects this fraction  --
 accuracies cannot be maintained if update messages never arrive.
In contrast,
 the thresholded performance measure ${\hat{\Aknow}}_{\upT}$
 remains high
 as ``good links'' can easily persist, 
 although in fig. \ref{fig-results-back-cfar}
 we can see this drop off for higher losses.
A key feature showing differences here 
 is the normalised risk measure $\bar{R}$.
For the Random, Timer, and Sequence tactics
 it is the same (at $\sim 1$), 
 as expected since these tactics
 all send messages at the same average rate, 
 regardless of efficiencies or accuracies.
In contrast,  
 the Filtered tactics exhibit much reduced risks, 
 although the variants with extra ``top up'' messages 
 are more risky, 
 as would be expected.

On fig. \ref{fig-results-back-cfar} we see that the 
 ``all links'' performance measure $\hat{\Aknow}$
 is uniformly low, 
 being approximately the same as the link probability 
 used to generate the ${\Elink}_{ij}$; 
 which implies that efficient links-- 
 as might be expected -- 
 lead to agents having high accuracy ${\Aknow}$ values on those links.
This is confirmed by the thresholded performance measure
 $\hat{\Aknow}_{\upT}$, 
 which remains high, 
 because it depends only on accurate belief values.

\begin{figure}
\begin{center}
\includegraphics[width=0.70\columnwidth]{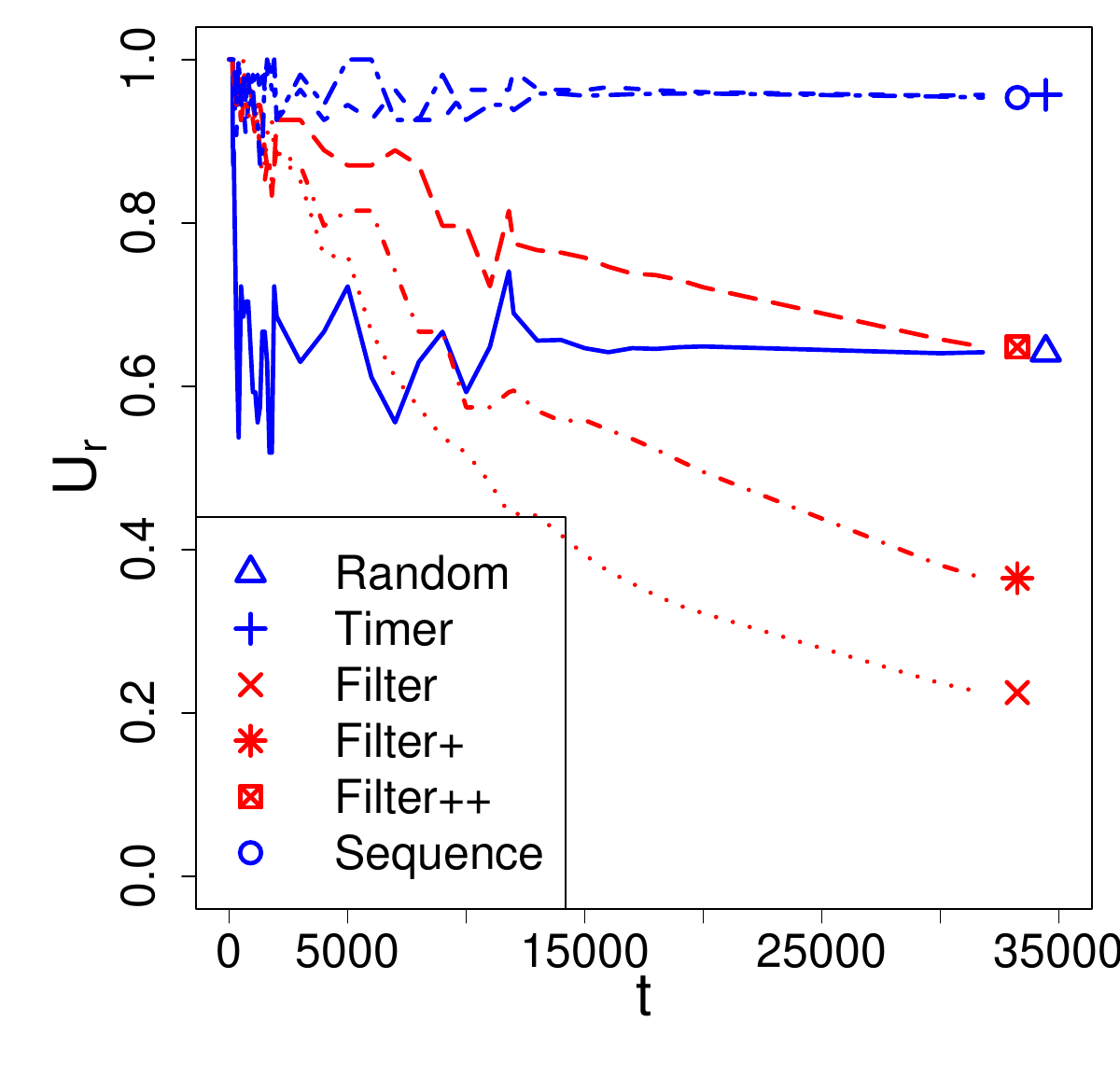}
\end{center}
\caption{
Swarm-averaged relative link counts $U_r$ as a function of simulation tick-count $t$
 for the different messaging tactics.
$U_r$ is the number of found accurate links
  (as determined by ${\Aknow}^i_j > \pthreshold$), 
 divided by the total number of efficient links 
  (as determined from ${\Elink}_{ij} > \pthreshold$).
We see that Random, Timer, and Sequential
 find steady state values, 
 but as time passes the Filtered tactics, 
 if not stopped, 
 lose the use of more and more links.
These results are for a relatively high loss of $r=\bar{\gamma}=\gamma \tau \nticks= 0.20$, 
 selected to emphasise the different behaviors, 
 but are much less extreme for smaller $\bar{\gamma}$ values.  
Note that $\bar{\gamma}$ is the counterpart of $r$
 in the continuum model.
}
\label{fig-results-back-utime}
\end{figure}

It should be noted again that although a useful steady state limit
 for the Random, Timer, and Sequence tactics is possible, 
 in the long term the Filtered tactics will all gradually 
 lose even good links to bad luck, 
 as seen on  fig. \ref{fig-results-back-utime}.
Thus on fig. \ref{fig-results-back-cfar},
 the Filtered-type results are only indicative of a medium term performance.
Nevertheless, 
 the beneficial risk minimisation of these Filtered tactics
 can be maintained, 
 e.g. by simply stopping the removal of filtered links 
 after some suitable multiple of $K$ ticks, 
 or perhaps setting a minimum number of links
 for any agent\footnote{These two suggestions, 
  whilst plausible, 
  and quite workable rules-of-thumb,
  can still fail.}.

%
\section{Discussion}\label{S-discussion}

The preceeding results 
 have shown simple scoping and initial results 
 for our scenario considering 
 agent-swarm management under stringent communications constraint.
In particular, 
 we have:

\begin{enumerate}

\item
 developed a simple agent information model
  allowing both a continuum and discrete communications paradigm;
  as well as a way of selecting some inter-agent communications links 
  that are preferential (more efficient) for use than others, 
  thus mimicking the effect of a spatial distribution of agents
  without extra complications.

\item
 discussed the need for a distinction between 
 the contents of the model
 (in particular ${\Aknow}^i_j$, $\gamma$, ${\Elink}_{ij}$)
 and what quantities an agent $a$ should be allowed to use when taking action
 (e.g. not any of ${\Elink}_{ij}$).

\item
 presented some metrics, 
 notably those related to communications performance 
 (i.e. $\hat{\Aknow}$ and $\hat{\Aknow}_{\upT}$)
 and risk ($\hat{R}$)
 that can be used to evaluate both aspects of this
 performance-under-contraint scenario.

\item
 proposed a simple ``round trip time '' test that agents can use
 to deprecate messaging over inefficient links ${\Elink}_{ij}$, 
 and which relies only on explicitly agent-knowable timing data 
 (i.e. the Filter, Filter+, and Filter++ communications tactics).

\end{enumerate}

We showed that the Filter tactics did reduce the usage 
 of inefficent links without significantly affecting performance, 
 at least in the medium term,
 and that simple modification (i.e. Filter+ and Filter++)
 can greatly moderate the long time failure of the unmodifed Filter tactic.
{Although not addressed in this initial study, 
 which is only intended to introduce the problem, 
 modifications to these tactics could
 not only stop the long term failure,}
 but would e.g. enhance responsiveness to changes in the 
 environment ${\Elink}_{ij}$.
Nevertheless, 
 this double-edge behaviour from the Filter tactics 
 does act a salutary warning 
 {that such tactics 
 need to be checked for possible failure modes
 as well as for their adaptability.}

It is important to note that 
 here we have attempted as far as possible 
 to use performance measures
 that are calculated in ways that 
 only depend on agent properties, 
 and then preferably on properties that an agent is \emph{aware} of,
 in combination with plausible threshold parameters.
{This is deliberate, 
 because agents need metrics based only 
 on what they are aware of;
 and
 since the (e.g.) ${\Elink_{ij}}$ are explicitly unknowns, 
 they should not be used.
Nevertheless, 
 because our information model is so abstract, 
 we can only judge whether an agent thinks the $a \leftrightarrow b$ link
 is accurate by using ${\Aknow}^a_b$ regardless.
However, 
 we can still assume the results of such shortcuts based on ${\Aknow}^a_b$ values
 are somehow representative of a comparable judgement
 an agent might make
 based on an actual basket of data containing locations and uncertainties.}
Only in fig. \ref{fig-results-back-utime}
 do we have recourse to non-agent properties, 
 but that is simply a normalization 
 based on thresholded ${\Elink_{ij}}$ values, 
 to estimate to what extent the swarm 
 might have accurate information 
 about the efficient links that are present.

%
\section{Conclusion}\label{S-conclusion}

Here we have considered an abstract implementation
 of agent-swarm behaviour 
 in terms of a mathematical multi-agent model.
We presented a simple continuum communications model which 
 allowed some analytic treatment, 
 but mainly focussed on a discrete communications
 version with a stochastic implementation
 which more accurately reflected any likely implementation.
In particular, 
 the stochastic version also provides us with information about outcome distributions
 as well as average- behaviour;
 something the continuum model does not generate.

We can see how these distributions are key to understanding 
 the success/failure threshold of the swarm
 as seen in figs. \ref{fig-results-flat-simple-hh}
 and \ref{fig-results-back-simple-hh}.
Whilst in the clear-success and clear-failure regimes
 (i.e. either low-loss or high-loss)
 the distribution of information available to the agents 
 is indeed narrowly peaked, 
 we see that in the transition region it broadens asymmetrically.
This behaviour is key since we consider scenarios where we must operate our swarm
 on {minimal} communications, 
 and \emph{it is somewhere in this transition region} that we would
 aim to position our operating parameters --
 i.e. with enough messaging to ensure swarm coherence, 
 but only \emph{just} enough,
 so that risk of detection remains at a minimum.

Although the distributions in performance are key to a detailed understanding, 
 we also used summary metrics, 
 notably the swarm risk $\hat{R}$
 and the thresholded performance measure $\hat{\Aknow}_{\upT}$.
By thresholding, 
 we enable the metric to ignore intrinsically inefficient and unecessary links, 
 which should indeed be irrelevant to any practical performance measure.
Further, 
 as part of the thresholding we normalised on the basis of a 
 minimal ER network structure, 
 so that poorly linked agents were still penalised in the metric, 
 but in a way not dependent on any information specific to a particular 
 set of ${\Elink}_{ij}$.
One improvement that could be made here is to replace this 
 minimum ER criterion
 with one specific to spatially distributed agents, 
 rather than probabilistically linked ones.

Finally, 
 the model used here was highly abstracted and not strictly applicable
 to any realistic agent-swarm implementation.
{Nevertheless, 
 we have here proposed an interesting and novel scenario
 with distinctive features, 
 and presented results and conclusions to aid in its understanding.
Since the lessons exhibited here will provide useful background 
 for more complicated scenarios and implementations, 
 we have accordingly developed our simulation software further, 
 allowing more sophisticated modelling \cite{Kinsler-EHMW-2022anglen}.}

\vspace{6pt} 
%

\acknowledgments{The authors would like to thank
 the Newton Gateway of Mathematics and their 
 \href{https://gateway.newton.ac.uk/event/tgmw88}{``Mathematical 
Study Group for Electromagnetic Challenges''}
 event which initiated this project.
We also acknowledge invaluable discussions
 with Richard Claridge and Adam Todd of PA Consulting, 
 as well as Paul Howland and Louise Hazelton of DSTL.}

%
\input{Kinsler-HEMW-2022arxiv.bbl}

%

\section*{Appendix: The continuum model steady state}\label{S-appdx-css}

We can normalise the dynamical equation 
 by scaling with respect to $\gamma$, 
 so that with $s = \gamma t$, 
 we have a dynamical equation
 for any accuracy, 
 which is
~
\begin{align}
  \frac{d}{ds}
  {\Aknow}^a_b
&=
 -
  \left( {\Aknow}^a_b - {\Akmin} \right)
 +
  \frac{ {\Trate}^b_a
         {\Elink}_{ba}
       }
       {\gamma}
  {\Aknow}^b_a
   \left(1-{\Aknow}^a_b\right)
\\
&=
 -
  \left( {\Aknow}^a_b - {\Akmin} \right)
 +
  \frac{1}{R_{ab}}
  {\Aknow}^b_a
   \left(1-{\Aknow}^a_b\right)
,
\end{align}
 where $R_{ab} = \gamma / {\Trate}^b_a {\Elink}_{ba}$.

Thus the steady state un-normalised (\& normalised) accuracy ${\Aknow}^a_b$ is
~
\begin{align}
  {\Aknow}^a_b
&=
  \frac{  \gamma {\Akmin}  +  {\Trate}^b_a {\Elink}_{ba} {\Aknow}^b_a  }
       {  \gamma           +  {\Trate}^b_a {\Elink}_{ba} {\Aknow}^b_a  }
\quad 
=
  \frac{  {\Akmin}  +  R_{ab}^{-1} {\Aknow}^b_a  }
       {  1         +  R_{ab}^{-1}  {\Aknow}^b_a  }
,
\end{align}
 and we can substitute either of these expressions 
 into itself (with with reversed $a,b$ indices),
 to get a polynomial for ${\Aknow}^a_b$.

We create a forward loss rate $r = \gamma /{\Trate}^b_a {\Elink}_{ba}$
 and a backward rate $r' = \gamma /{\Trate}^a_b {\Elink}_{ab}$, 
 so that 
 we have
~
\begin{align}
  {\Aknow}^a_b
&=
  \frac{  r {\Akmin}  +  {\Aknow}^b_a  }
       {  r           +  {\Aknow}^b_a  }
\\
&=
  \frac{  r {\Akmin}
         +  
          \frac{  r' {\Akmin}  +  {\Aknow}^a_b  }
               {  r'           +  {\Aknow}^a_b  }
       }
       {  r
         +  
          \frac{  r' {\Akmin}  +  {\Aknow}^a_b  }
               {  r'           +  {\Aknow}^a_b  }
       }
\\
  {\Aknow}^a_b
  \left\{
       {  r
          \left[ r'           +  {\Aknow}^a_b \right]
         +  
          \left[ r' {\Akmin}  +  {\Aknow}^a_b \right]
      }
  \right\}
&=
          r {\Akmin}
          \left[ r'           +  {\Aknow}^a_b \right]
         +
          \left[ r' {\Akmin}  +  {\Aknow}^a_b \right]
\\
  r'
  \left[
    r
   +
    {\Akmin}
  \right]
  {\Aknow}^a_b
 +
  \left[
    r
   +
    1
  \right]
  \left[
    {\Aknow}^a_b
  \right]^2
&=
  r' {\Akmin}
  \left[
    r
   +
    1
  \right]
 +
   \left[
     r {\Akmin} + 1
   \right]
   {\Aknow}^a_b
\\
  \left[
    r' r
   +
    \left( r' - r \right) {\Akmin}
  \right]
  {\Aknow}^a_b
 +
  \left[
    r
   +
    1
  \right]
  \left[
    {\Aknow}^a_b
  \right]^2
&=
  r'
 {\Akmin}
  \left[
    r
   +
    1
  \right]
 +
   {\Aknow}^a_b
.
\end{align}

Hence
~
\begin{align}
  \left[
    r
   +
    1
  \right]
  \left[
    {\Aknow}^a_b
  \right]^2
 +
  \left[
    r' r
   +
    \left( r' - r \right) {\Akmin}
   -
    1
  \right]
  {\Aknow}^a_b
 -
  r'
 {\Akmin}
  \left[
    r
   +
    1
  \right]
&=
  0
\end{align}

And if $r=r'$, 
 which would be reasonable for a symmetric environment ${\Elink}_{ab}={\Elink}_{ba}$,
 where both agent $a$ and $b$ were transmitting at the same default rate, 
 we have the simpler form
~
\begin{align}
  \left[
    {\Aknow}^a_b
  \right]^2
 +
  \left[
    r
   -
    1
  \right]
  {\Aknow}^a_b
 -
  r {\Akmin}
&=
  0
,
\label{eqn-appdx-cns-quadratic}
\end{align}
 an expression which 
 has just two free parameters, 
 the rate ratio $r$
 and the minimum information ${\Akmin}$.

Thus
~
\begin{align}
    {\Aknow}^a_b
&=
  \frac{1-r}{2}
 \pm
  \frac{1}{2}
  \sqrt{ r^2 - 2r +1 - 4(-r).{\Akmin} }
,
\\
&=
  \frac{1}{2}
 -
  \frac{r}{2}
 \pm
  \frac{1}{2}
  \sqrt{ r^2 - 2 r \left( 1 - 2 {\Akmin} \right)  + 1 }
.
\end{align}

If ${\Akmin}=0$, 
 then
~
\begin{align}
    {\Aknow}^a_b
&= 
  \frac{1}{2}
 -
  \frac{r}{2}
 \pm
  \frac{1}{2}
  \left( r - 1 \right)
,
\end{align}
 which has two solutions; 
 firstly the 
  zero accuracy case ${\Aknow}^a_b=0$,
 and secondly the 
  finite-accuracy case ${\Aknow}^a_b = 1-r$.
However, 
 if ${\Akmin}>0$, 
 the ``zero accuracy'' solution (sign choice ``$+$'') 
 is pushed negative so that 
 only the finite-accuracy one remains.

%

\end{document}

%% file: fig-agentio-b.pdf_t
\begin{picture}(0,0)%
\includegraphics{fig-agentio-b.pdf}%
\end{picture}%
\setlength{\unitlength}{704sp}%
\begingroup\makeatletter\ifx\SetFigFont\undefined%
\gdef\SetFigFont#1#2#3#4#5{%
  \reset@font\fontsize{#1}{#2pt}%
  \fontfamily{#3}\fontseries{#4}\fontshape{#5}%
  \selectfont}%
\fi\endgroup%
\begin{picture}(11728,9388)(617,-8715)
\put(1290,-4023){\makebox(0,0)[lb]{\smash{{\SetFigFont{5}{6.0}{\rmdefault}{\mddefault}{\updefault}{\color[rgb]{0,0,0}$\Phi^a_i$}%
}}}}
\put(1470,-4833){\makebox(0,0)[lb]{\smash{{\SetFigFont{5}{6.0}{\rmdefault}{\mddefault}{\updefault}{\color[rgb]{0,0,0}$\alpha^a_i$}%
}}}}
\put(10703,-191){\makebox(0,0)[lb]{\smash{{\SetFigFont{5}{6.0}{\rmdefault}{\mddefault}{\updefault}{\color[rgb]{0,0,0}$\Phi^b_i$}%
}}}}
\put(10928,-956){\makebox(0,0)[lb]{\smash{{\SetFigFont{5}{6.0}{\rmdefault}{\mddefault}{\updefault}{\color[rgb]{0,0,0}$\alpha^b_i$}%
}}}}
\put(8821,-7218){\makebox(0,0)[lb]{\smash{{\SetFigFont{5}{6.0}{\rmdefault}{\mddefault}{\updefault}{\color[rgb]{0,0,0}$\Phi^c_i$}%
}}}}
\put(9001,-8028){\makebox(0,0)[lb]{\smash{{\SetFigFont{5}{6.0}{\rmdefault}{\mddefault}{\updefault}{\color[rgb]{0,0,0}$\alpha^c_i$}%
}}}}
\put(8596,-1816){\rotatebox{22.0}{\makebox(0,0)[lb]{\smash{{\SetFigFont{5}{6.0}{\rmdefault}{\mddefault}{\updefault}{\color[rgb]{0,0,0}$L_{ba} \simeq 0$}%
}}}}}
\put(4321,-2716){\rotatebox{20.0}{\makebox(0,0)[rb]{\smash{{\SetFigFont{5}{6.0}{\rmdefault}{\mddefault}{\updefault}{\color[rgb]{0,0,0}$L_{ab} \simeq 0$}%
}}}}}
\put(5806,-5461){\rotatebox{338.0}{\makebox(0,0)[lb]{\smash{{\SetFigFont{5}{6.0}{\rmdefault}{\mddefault}{\updefault}{\color[rgb]{0,0,0}$L_{ac} \simeq 1$}%
}}}}}
\put(5626,-6721){\rotatebox{338.0}{\makebox(0,0)[rb]{\smash{{\SetFigFont{5}{6.0}{\rmdefault}{\mddefault}{\updefault}{\color[rgb]{0,0,0}$L_{ca} \simeq 1$}%
}}}}}
\put(9991,-2941){\rotatebox{74.0}{\makebox(0,0)[rb]{\smash{{\SetFigFont{5}{6.0}{\rmdefault}{\mddefault}{\updefault}{\color[rgb]{0,0,0}$L_{bc} \simeq 1$}%
}}}}}
\put(10576,-5236){\rotatebox{76.0}{\makebox(0,0)[lb]{\smash{{\SetFigFont{5}{6.0}{\rmdefault}{\mddefault}{\updefault}{\color[rgb]{0,0,0}$L_{bc} \simeq 1$}%
}}}}}
\end{picture}%

%% file: fig-venndiagram.pdf_t
\begin{picture}(0,0)%
\includegraphics{fig-venndiagram.pdf}%
\end{picture}%
\setlength{\unitlength}{704sp}%
\begingroup\makeatletter\ifx\SetFigFont\undefined%
\gdef\SetFigFont#1#2#3#4#5{%
  \reset@font\fontsize{#1}{#2pt}%
  \fontfamily{#3}\fontseries{#4}\fontshape{#5}%
  \selectfont}%
\fi\endgroup%
\begin{picture}(10625,7563)(417,-7185)
\put(8642,-4934){\makebox(0,0)[b]{\smash{{\SetFigFont{5}{6.0}{\rmdefault}{\mddefault}{\updefault}{\color[rgb]{0,0,0}$\Phi^b_j$}%
}}}}
\put(5878,-5306){\makebox(0,0)[b]{\smash{{\SetFigFont{5}{6.0}{\rmdefault}{\mddefault}{\updefault}{\color[rgb]{0,0,0}$\alpha^b_i$}%
}}}}
\put(8418,-2246){\makebox(0,0)[b]{\smash{{\SetFigFont{5}{6.0}{\rmdefault}{\mddefault}{\updefault}{\color[rgb]{0,0,0}$\Phi^a_j$}%
}}}}
\put(5653,-2496){\makebox(0,0)[b]{\smash{{\SetFigFont{5}{6.0}{\rmdefault}{\mddefault}{\updefault}{\color[rgb]{0,0,0}$\alpha^a_i$}%
}}}}
\put(2817,-4338){\makebox(0,0)[b]{\smash{{\SetFigFont{5}{6.0}{\rmdefault}{\mddefault}{\updefault}{\color[rgb]{0,0,0}$L_{ij}$}%
}}}}
\end{picture}%

%% file: fig-cnsquadwrapper.pdf_t
\begin{picture}(0,0)%
\includegraphics{fig-cnsquadwrapper.pdf}%
\end{picture}%
\setlength{\unitlength}{704sp}%
\begingroup\makeatletter\ifx\SetFigFont\undefined%
\gdef\SetFigFont#1#2#3#4#5{%
  \reset@font\fontsize{#1}{#2pt}%
  \fontfamily{#3}\fontseries{#4}\fontshape{#5}%
  \selectfont}%
\fi\endgroup%
\begin{picture}(9452,9182)(450,-8792)
\put(856,-4336){\rotatebox{90.0}{\makebox(0,0)[lb]{\smash{{\SetFigFont{5}{6.0}{\rmdefault}{\mddefault}{\updefault}{\color[rgb]{0,0,0}$\Akmin$}%
}}}}}
\put(5356,-8206){\makebox(0,0)[lb]{\smash{{\SetFigFont{5}{6.0}{\rmdefault}{\mddefault}{\updefault}{\color[rgb]{0,0,0}$r$}%
}}}}
\end{picture}%

%% file: fig-Flat-S0-200-hhslice-picc.pdf_t
\begin{picture}(0,0)%
\includegraphics{fig-Flat-S0-200-hhslice-picc.pdf}%
\end{picture}%
\setlength{\unitlength}{2901sp}%
\begingroup\makeatletter\ifx\SetFigFont\undefined%
\gdef\SetFigFont#1#2#3#4#5{%
  \reset@font\fontsize{#1}{#2pt}%
  \fontfamily{#3}\fontseries{#4}\fontshape{#5}%
  \selectfont}%
\fi\endgroup%
\begin{picture}(6176,6274)(4,-5660)
\put(3376,-5551){\makebox(0,0)[b]{\smash{{\SetFigFont{14}{16.8}{\rmdefault}{\mddefault}{\updefault}{\color[rgb]{0,0,0}\huge{$\bar{\gamma} =  \gamma \tau K$}}%
}}}}
\put(271,-2041){\rotatebox{90.0}{\makebox(0,0)[b]{\smash{{\SetFigFont{14}{16.8}{\rmdefault}{\mddefault}{\updefault}{\color[rgb]{0,0,0}\huge{$\Phi$}}%
}}}}}
\end{picture}%

%% file: fig-Flat-S1-200-hhslice-picc.pdf_t
\begin{picture}(0,0)%
\includegraphics{fig-Flat-S1-200-hhslice-picc.pdf}%
\end{picture}%
\setlength{\unitlength}{2901sp}%
\begingroup\makeatletter\ifx\SetFigFont\undefined%
\gdef\SetFigFont#1#2#3#4#5{%
  \reset@font\fontsize{#1}{#2pt}%
  \fontfamily{#3}\fontseries{#4}\fontshape{#5}%
  \selectfont}%
\fi\endgroup%
\begin{picture}(6176,6274)(4,-5660)
\put(3376,-5551){\makebox(0,0)[b]{\smash{{\SetFigFont{14}{16.8}{\rmdefault}{\mddefault}{\updefault}{\color[rgb]{0,0,0}\huge{$\bar{\gamma} =  \gamma \tau K$}}%
}}}}
\put(271,-2041){\rotatebox{90.0}{\makebox(0,0)[b]{\smash{{\SetFigFont{14}{16.8}{\rmdefault}{\mddefault}{\updefault}{\color[rgb]{0,0,0}\huge{$\Phi$}}%
}}}}}
\end{picture}%

%% file: fig-Back-Sr0-020-hhslice-picc.pdf_t
\begin{picture}(0,0)%
\includegraphics{fig-Back-Sr0-020-hhslice-picc.pdf}%
\end{picture}%
\setlength{\unitlength}{2901sp}%
\begingroup\makeatletter\ifx\SetFigFont\undefined%
\gdef\SetFigFont#1#2#3#4#5{%
  \reset@font\fontsize{#1}{#2pt}%
  \fontfamily{#3}\fontseries{#4}\fontshape{#5}%
  \selectfont}%
\fi\endgroup%
\begin{picture}(6176,6274)(4,-5660)
\put(3376,-5551){\makebox(0,0)[b]{\smash{{\SetFigFont{14}{16.8}{\rmdefault}{\mddefault}{\updefault}{\color[rgb]{0,0,0}\huge{$\bar{\gamma} =  \gamma \tau K$}}%
}}}}
\put(271,-2041){\rotatebox{90.0}{\makebox(0,0)[b]{\smash{{\SetFigFont{14}{16.8}{\rmdefault}{\mddefault}{\updefault}{\color[rgb]{0,0,0}\huge{$\Phi$}}%
}}}}}
\end{picture}%

%% file: fig-Back-Sr2-020-hhslice-picc.pdf_t
\begin{picture}(0,0)%
\includegraphics{fig-Back-Sr2-020-hhslice-picc.pdf}%
\end{picture}%
\setlength{\unitlength}{2901sp}%
\begingroup\makeatletter\ifx\SetFigFont\undefined%
\gdef\SetFigFont#1#2#3#4#5{%
  \reset@font\fontsize{#1}{#2pt}%
  \fontfamily{#3}\fontseries{#4}\fontshape{#5}%
  \selectfont}%
\fi\endgroup%
\begin{picture}(6176,6274)(4,-5660)
\put(3376,-5551){\makebox(0,0)[b]{\smash{{\SetFigFont{14}{16.8}{\rmdefault}{\mddefault}{\updefault}{\color[rgb]{0,0,0}\huge{$\bar{\gamma} =  \gamma \tau K$}}%
}}}}
\put(271,-2041){\rotatebox{90.0}{\makebox(0,0)[b]{\smash{{\SetFigFont{14}{16.8}{\rmdefault}{\mddefault}{\updefault}{\color[rgb]{0,0,0}\huge{$\Phi$}}%
}}}}}
\end{picture}%

%% file: Kinsler-HEMW-2022arxiv.bbl
%

%% file: Kinsler-HEMW-2022arxiv.bbl
\begin{thebibliography}{33}%
\makeatletter
\providecommand \@ifxundefined [1]{%
 \@ifx{#1\undefined}
}%
\providecommand \@ifnum [1]{%
 \ifnum #1\expandafter \@firstoftwo
 \else \expandafter \@secondoftwo
 \fi
}%
\providecommand \@ifx [1]{%
 \ifx #1\expandafter \@firstoftwo
 \else \expandafter \@secondoftwo
 \fi
}%
\providecommand \natexlab [1]{#1}%
\providecommand \enquote  [1]{``#1''}%
\providecommand \bibnamefont  [1]{#1}%
\providecommand \bibfnamefont [1]{#1}%
\providecommand \citenamefont [1]{#1}%
\providecommand \href@noop [0]{\@secondoftwo}%
\providecommand \href [0]{\begingroup \@sanitize@url \@href}%
\providecommand \@href[1]{\@@startlink{#1}\@@href}%
\providecommand \@@href[1]{\endgroup#1\@@endlink}%
\providecommand \@sanitize@url [0]{\catcode `\\12\catcode `\$12\catcode
  `\&12\catcode `\#12\catcode `\^12\catcode `\_12\catcode `\%12\relax}%
\providecommand \@@startlink[1]{}%
\providecommand \@@endlink[0]{}%
\providecommand \url  [0]{\begingroup\@sanitize@url \@url }%
\providecommand \@url [1]{\endgroup\@href {#1}{\urlprefix }}%
\providecommand \urlprefix  [0]{URL }%
\providecommand \Eprint [0]{\href }%
\providecommand \doibase [0]{https://doi.org/}%
\providecommand \selectlanguage [0]{\@gobble}%
\providecommand \bibinfo  [0]{\@secondoftwo}%
\providecommand \bibfield  [0]{\@secondoftwo}%
\providecommand \translation [1]{[#1]}%
\providecommand \BibitemOpen [0]{}%
\providecommand \bibitemStop [0]{}%
\providecommand \bibitemNoStop [0]{.\EOS\space}%
\providecommand \EOS [0]{\spacefactor3000\relax}%
\providecommand \BibitemShut  [1]{\csname bibitem#1\endcsname}%
\let\auto@bib@innerbib\@empty
\bibitem [{\citenamefont {Bekmezcia}\ \emph {et~al.}(2013)\citenamefont
  {Bekmezcia}, \citenamefont {Sahingoza},\ and\ \citenamefont
  {Temel}}]{Bekmezcia-ST-2013ahn}%
  \BibitemOpen
  \bibfield  {author} {\bibinfo {author} {\bibfnamefont {I.}~\bibnamefont
  {Bekmezcia}}, \bibinfo {author} {\bibfnamefont {O.~K.}\ \bibnamefont
  {Sahingoza}},\ and\ \bibinfo {author} {\bibfnamefont {S.}~\bibnamefont
  {Temel}},\ }\bibfield  {title} {\bibinfo {title} {Flying ad-hoc networks
  (FANETs): A survey},\ }\href {https://doi.org/10.1016/j.adhoc.2012.12.004}
  {\bibfield  {journal} {\bibinfo  {journal} {Ad Hoc Networks}\ }\textbf
  {\bibinfo {volume} {11}},\ \bibinfo {pages} {1254} (\bibinfo {year}
  {2013})}\BibitemShut {NoStop}%
\bibitem [{\citenamefont {Shakhatreh}\ \emph {et~al.}(2019)\citenamefont
  {Shakhatreh}, \citenamefont {Sawalmeh}, \citenamefont {Al-Fuqaha},
  \citenamefont {Dou}, \citenamefont {Almaita}, \citenamefont {Khalil},
  \citenamefont {Othman}, \citenamefont {Khreishah},\ and\ \citenamefont
  {Guizani}}]{Shakhatreh-SFDAKOKG-2019ieeea}%
  \BibitemOpen
  \bibfield  {author} {\bibinfo {author} {\bibfnamefont {H.}~\bibnamefont
  {Shakhatreh}}, \bibinfo {author} {\bibfnamefont {A.~H.}\ \bibnamefont
  {Sawalmeh}}, \bibinfo {author} {\bibfnamefont {A.}~\bibnamefont {Al-Fuqaha}},
  \bibinfo {author} {\bibfnamefont {Z.}~\bibnamefont {Dou}}, \bibinfo {author}
  {\bibfnamefont {E.}~\bibnamefont {Almaita}}, \bibinfo {author} {\bibfnamefont
  {I.}~\bibnamefont {Khalil}}, \bibinfo {author} {\bibfnamefont {N.~S.}\
  \bibnamefont {Othman}}, \bibinfo {author} {\bibfnamefont {A.}~\bibnamefont
  {Khreishah}},\ and\ \bibinfo {author} {\bibfnamefont {M.}~\bibnamefont
  {Guizani}},\ }\bibfield  {title} {\bibinfo {title} {Unmanned aerial vehicles
  (UAVs): A survey on civil applications and key research challenges},\ }\href
  {https://doi.org/10.1109/ACCESS.2019.2909530} {\bibfield  {journal} {\bibinfo
   {journal} {IEEE Access}\ }\textbf {\bibinfo {volume} {7}},\ \bibinfo {pages}
  {48572} (\bibinfo {year} {2019})}\BibitemShut {NoStop}%
\bibitem [{\citenamefont {Haider}\ \emph {et~al.}(2022)\citenamefont {Haider},
  \citenamefont {Nauman}, \citenamefont {Jamshed}, \citenamefont {Jiang},
  \citenamefont {Batool},\ and\ \citenamefont {Kim}}]{Haider-NJJBK-2022mdpi}%
  \BibitemOpen
  \bibfield  {author} {\bibinfo {author} {\bibfnamefont {S.~K.}\ \bibnamefont
  {Haider}}, \bibinfo {author} {\bibfnamefont {A.}~\bibnamefont {Nauman}},
  \bibinfo {author} {\bibfnamefont {M.~A.}\ \bibnamefont {Jamshed}}, \bibinfo
  {author} {\bibfnamefont {A.}~\bibnamefont {Jiang}}, \bibinfo {author}
  {\bibfnamefont {S.}~\bibnamefont {Batool}},\ and\ \bibinfo {author}
  {\bibfnamefont {S.~W.}\ \bibnamefont {Kim}},\ }\bibfield  {title} {\bibinfo
  {title} {Internet of drones: Routing algorithms, techniques and challenges},\
  }\href {https://doi.org/10.3390/math10091488} {\bibfield  {journal} {\bibinfo
   {journal} {Mathematics}\ }\textbf {\bibinfo {volume} {10}},\ \bibinfo
  {pages} {1488} (\bibinfo {year} {2022})}\BibitemShut {NoStop}%
\bibitem [{\citenamefont {Rejeb}\ \emph {et~al.}(2022)\citenamefont {Rejeb},
  \citenamefont {Abdollahi}, \citenamefont {Rejeb},\ and\ \citenamefont
  {Treiblmaier}}]{Rejeb-ART-2022cea}%
  \BibitemOpen
  \bibfield  {author} {\bibinfo {author} {\bibfnamefont {A.}~\bibnamefont
  {Rejeb}}, \bibinfo {author} {\bibfnamefont {A.}~\bibnamefont {Abdollahi}},
  \bibinfo {author} {\bibfnamefont {K.}~\bibnamefont {Rejeb}},\ and\ \bibinfo
  {author} {\bibfnamefont {H.}~\bibnamefont {Treiblmaier}},\ }\bibfield
  {title} {\bibinfo {title} {Drones in agriculture: A review and bibliometric
  analysis},\ }\href {https://doi.org/10.1016/j.compag.2022.107017} {\bibfield
  {journal} {\bibinfo  {journal} {Computers and Electronics in Agriculture}\
  }\textbf {\bibinfo {volume} {198}},\ \bibinfo {pages} {107017} (\bibinfo
  {year} {2022})}\BibitemShut {NoStop}%
\bibitem [{\citenamefont {Benarbia}\ and\ \citenamefont
  {Kyamakya}(2022)}]{Benarbia-K-2022mdpi}%
  \BibitemOpen
  \bibfield  {author} {\bibinfo {author} {\bibfnamefont {T.}~\bibnamefont
  {Benarbia}}\ and\ \bibinfo {author} {\bibfnamefont {K.}~\bibnamefont
  {Kyamakya}},\ }\bibfield  {title} {\bibinfo {title} {A literature review of
  drone-based package delivery logistics systems and their implementation
  feasibility},\ }\href {https://doi.org/10.3390/su14010360} {\bibfield
  {journal} {\bibinfo  {journal} {Sustainability}\ }\textbf {\bibinfo {volume}
  {14}},\ \bibinfo {pages} {360} (\bibinfo {year} {2022})}\BibitemShut
  {NoStop}%
\bibitem [{\citenamefont {Kucharczyk}\ and\ \citenamefont
  {Hugenholtz}(2021)}]{Kucharczyk-H-2021rse}%
  \BibitemOpen
  \bibfield  {author} {\bibinfo {author} {\bibfnamefont {M.}~\bibnamefont
  {Kucharczyk}}\ and\ \bibinfo {author} {\bibfnamefont {C.~H.}\ \bibnamefont
  {Hugenholtz}},\ }\bibfield  {title} {\bibinfo {title} {Remote sensing of
  natural hazard-related disasters with small drones: Global trends, biases,
  and research opportunities},\ }\href
  {https://doi.org/10.1016/j.rse.2021.112577} {\bibfield  {journal} {\bibinfo
  {journal} {Remote Sensing of Environment}\ }\textbf {\bibinfo {volume}
  {264}},\ \bibinfo {pages} {112577} (\bibinfo {year} {2021})}\BibitemShut
  {NoStop}%
\bibitem [{\citenamefont {Stodola}\ \emph {et~al.}(2022)\citenamefont
  {Stodola}, \citenamefont {Nohel}, \citenamefont {Fagiolini}, \citenamefont
  {Vasik}, \citenamefont {Turi}, \citenamefont {Bruzzone}, \citenamefont
  {Pickl}, \citenamefont {Neumann},\ and\ \citenamefont
  {Stodola}}]{Stodola-NFVTBPNS-2022mesa}%
  \BibitemOpen
  \bibfield  {author} {\bibinfo {author} {\bibfnamefont {P.}~\bibnamefont
  {Stodola}}, \bibinfo {author} {\bibfnamefont {J.}~\bibnamefont {Nohel}},
  \bibinfo {author} {\bibfnamefont {A.}~\bibnamefont {Fagiolini}}, \bibinfo
  {author} {\bibfnamefont {P.}~\bibnamefont {Vasik}}, \bibinfo {author}
  {\bibfnamefont {M.}~\bibnamefont {Turi}}, \bibinfo {author} {\bibfnamefont
  {A.}~\bibnamefont {Bruzzone}}, \bibinfo {author} {\bibfnamefont
  {S.}~\bibnamefont {Pickl}}, \bibinfo {author} {\bibfnamefont
  {V.}~\bibnamefont {Neumann}},\ and\ \bibinfo {author} {\bibfnamefont
  {P.}~\bibnamefont {Stodola}},\ }\bibfield  {title} {\bibinfo {title}
  {Reconnaissance in complex environment with no-fly zones using a swarm of
  unmanned aerial vehicles},\ }in\ \href
  {https://doi.org/10.1007/978-3-030-98260-7_19} {\emph {\bibinfo {booktitle}
  {Modelling and Simulation for Autonomous Systems}}},\ \bibinfo {series and
  number} {Lecture Notes in Computer Science}\ (\bibinfo  {publisher} {Springer
  International Publishing},\ \bibinfo {year} {2022})\ pp.\ \bibinfo {pages}
  {308--321}\BibitemShut {NoStop}%
\bibitem [{\citenamefont {Maza}\ \emph {et~al.}(2011)\citenamefont {Maza},
  \citenamefont {Caballero}, \citenamefont {Capit{\'a}n}, \citenamefont
  {Mart{\'{\i}}nez-De-Dios},\ and\ \citenamefont
  {Ollero}}]{Maza-CCMO-2011jirs}%
  \BibitemOpen
  \bibfield  {author} {\bibinfo {author} {\bibfnamefont {I.}~\bibnamefont
  {Maza}}, \bibinfo {author} {\bibfnamefont {F.}~\bibnamefont {Caballero}},
  \bibinfo {author} {\bibfnamefont {J.}~\bibnamefont {Capit{\'a}n}}, \bibinfo
  {author} {\bibfnamefont {J.~R.}\ \bibnamefont {Mart{\'{\i}}nez-De-Dios}},\
  and\ \bibinfo {author} {\bibfnamefont {A.}~\bibnamefont {Ollero}},\
  }\bibfield  {title} {\bibinfo {title} {Experimental results in multi-uav
  coordination for disaster management and civil security applications},\
  }\href {https://doi.org/10.1007/s10846-010-9497-5} {\bibfield  {journal}
  {\bibinfo  {journal} {Journal of Intelligent and Robotic Systems}\ }\textbf
  {\bibinfo {volume} {61}},\ \bibinfo {pages} {563} (\bibinfo {year}
  {2011})}\BibitemShut {NoStop}%
\bibitem [{\citenamefont {Cameron}\ \emph {et~al.}(2010)\citenamefont
  {Cameron}, \citenamefont {Symington}, \citenamefont {Trigoni},\ and\
  \citenamefont {Waharte}}]{Cameron-SW-2010suaave}%
  \BibitemOpen
  \bibfield  {author} {\bibinfo {author} {\bibfnamefont {S.}~\bibnamefont
  {Cameron}}, \bibinfo {author} {\bibfnamefont {A.~C.}\ \bibnamefont
  {Symington}}, \bibinfo {author} {\bibfnamefont {N.}~\bibnamefont {Trigoni}},\
  and\ \bibinfo {author} {\bibfnamefont {S.}~\bibnamefont {Waharte}},\
  }\bibfield  {title} {\bibinfo {title} {Suaave: Combining aerial robots and
  wireless networking},\ }in\ \href@noop {} {\emph {\bibinfo {booktitle} {25th
  Bristol International UAV Systems Conference}}}\ (\bibinfo {year}
  {2010})\BibitemShut {NoStop}%
\bibitem [{\citenamefont {Waharte}\ \emph {et~al.}(2009)\citenamefont
  {Waharte}, \citenamefont {Trigoni},\ and\ \citenamefont
  {Julier}}]{Waharte-TJ-2009suaave}%
  \BibitemOpen
  \bibfield  {author} {\bibinfo {author} {\bibfnamefont {S.}~\bibnamefont
  {Waharte}}, \bibinfo {author} {\bibfnamefont {N.}~\bibnamefont {Trigoni}},\
  and\ \bibinfo {author} {\bibfnamefont {S.}~\bibnamefont {Julier}},\
  }\bibfield  {title} {\bibinfo {title} {Coordinated search with a swarm of
  uavs},\ }in\ \href {https://doi.org/10.1109/SAHCNW.2009.5172925} {\emph
  {\bibinfo {booktitle} {2009 6th IEEE Annual Communications Society Conference
  on Sensor, Mesh and Ad Hoc Communications and Networks Workshops}}}\
  (\bibinfo  {publisher} {IEEE},\ \bibinfo {year} {2009})\BibitemShut {NoStop}%
\bibitem [{\citenamefont {Horyna}\ \emph {et~al.}(2022)\citenamefont {Horyna},
  \citenamefont {Baca}, \citenamefont {Walter}, \citenamefont {Albani},
  \citenamefont {Hert}, \citenamefont {Ferrante},\ and\ \citenamefont
  {Saska}}]{Horyna-BWAHFS-2023ar}%
  \BibitemOpen
  \bibfield  {author} {\bibinfo {author} {\bibfnamefont {J.}~\bibnamefont
  {Horyna}}, \bibinfo {author} {\bibfnamefont {T.}~\bibnamefont {Baca}},
  \bibinfo {author} {\bibfnamefont {V.}~\bibnamefont {Walter}}, \bibinfo
  {author} {\bibfnamefont {D.}~\bibnamefont {Albani}}, \bibinfo {author}
  {\bibfnamefont {D.}~\bibnamefont {Hert}}, \bibinfo {author} {\bibfnamefont
  {E.}~\bibnamefont {Ferrante}},\ and\ \bibinfo {author} {\bibfnamefont
  {M.}~\bibnamefont {Saska}},\ }\bibfield  {title} {\bibinfo {title}
  {Decentralized swarms of unmanned aerial vehicles for search and rescue
  operations without explicit communication},\ }\href
  {https://doi.org/10.1007/s10514-022-10066-5} {\bibfield  {journal} {\bibinfo
  {journal} {Autonomous Robots}\ {\bibinfo {volume} {47}},\ \bibinfo {pages} {77--93}} (\bibinfo
  {year} {2023})}\BibitemShut {NoStop}%
\bibitem [{\citenamefont {Grimal}\ and\ \citenamefont
  {Sundaram}(2018)}]{Grimal-S-2019jcsl}%
  \BibitemOpen
  \bibfield  {author} {\bibinfo {author} {\bibfnamefont {F.}~\bibnamefont
  {Grimal}}\ and\ \bibinfo {author} {\bibfnamefont {J.}~\bibnamefont
  {Sundaram}},\ }\bibfield  {title} {\bibinfo {title} {Combat drones: Hives,
  swarms, and autonomous action?},\ }\href
  {https://doi.org/10.1093/jcsl/kry008} {\bibfield  {journal} {\bibinfo
  {journal} {Journal of Conflict and Security Law}\ }\textbf {\bibinfo {volume}
  {23}},\ \bibinfo {pages} {105} (\bibinfo {year} {2018})}\BibitemShut
  {NoStop}%
\bibitem [{\citenamefont {George}\ \emph {et~al.}(2011)\citenamefont {George},
  \citenamefont {Sujit},\ and\ \citenamefont {Sousa}}]{George-SS-2011jirs}%
  \BibitemOpen
  \bibfield  {author} {\bibinfo {author} {\bibfnamefont {J.}~\bibnamefont
  {George}}, \bibinfo {author} {\bibfnamefont {P.~B.}\ \bibnamefont {Sujit}},\
  and\ \bibinfo {author} {\bibfnamefont {J.~B.}\ \bibnamefont {Sousa}},\
  }\bibfield  {title} {\bibinfo {title} {Search strategies for multiple uav
  search and destroy missions},\ }\href
  {https://doi.org/10.1007/s10846-010-9486-8} {\bibfield  {journal} {\bibinfo
  {journal} {Journal of Intelligent and Robotic Systems}\ }\textbf {\bibinfo
  {volume} {61}},\ \bibinfo {pages} {355} (\bibinfo {year} {2011})}\BibitemShut
  {NoStop}%
\bibitem [{\citenamefont {Saeed}\ \emph {et~al.}(2022)\citenamefont {Saeed},
  \citenamefont {Omri}, \citenamefont {Abdel-Khalek}, \citenamefont {Ali},\
  and\ \citenamefont {Alotaibi}}]{Saeed-OAAA-2022}%
  \BibitemOpen
  \bibfield  {author} {\bibinfo {author} {\bibfnamefont {R.~A.}\ \bibnamefont
  {Saeed}}, \bibinfo {author} {\bibfnamefont {M.}~\bibnamefont {Omri}},
  \bibinfo {author} {\bibfnamefont {S.}~\bibnamefont {Abdel-Khalek}}, \bibinfo
  {author} {\bibfnamefont {E.~S.}\ \bibnamefont {Ali}},\ and\ \bibinfo {author}
  {\bibfnamefont {M.~F.}\ \bibnamefont {Alotaibi}},\ }\bibfield  {title}
  {\bibinfo {title} {Optimal path planning for drones based on swarm
  intelligence algorithm},\ }\href {https://doi.org/10.1007/s00521-022-06998-9}
  {\bibfield  {journal} {\bibinfo  {journal} {Neural Computing and
  Applications}\ }\textbf {\bibinfo {volume} {34}},\ \bibinfo {pages} {10133}
  (\bibinfo {year} {2022})}\BibitemShut {NoStop}%
\bibitem [{\citenamefont {Brambilla}\ \emph {et~al.}(2013)\citenamefont
  {Brambilla}, \citenamefont {Ferrante}, \citenamefont {Birattari},\ and\
  \citenamefont {Dorigo}}]{Brambilla-FBD-2013si}%
  \BibitemOpen
  \bibfield  {author} {\bibinfo {author} {\bibfnamefont {M.}~\bibnamefont
  {Brambilla}}, \bibinfo {author} {\bibfnamefont {E.}~\bibnamefont {Ferrante}},
  \bibinfo {author} {\bibfnamefont {M.}~\bibnamefont {Birattari}},\ and\
  \bibinfo {author} {\bibfnamefont {M.}~\bibnamefont {Dorigo}},\ }\bibfield
  {title} {\bibinfo {title} {Swarm robotics: a review from the swarm
  engineering perspective},\ }\href {https://doi.org/10.1007/s11721-012-0075-2}
  {\bibfield  {journal} {\bibinfo  {journal} {Swarm Intelligence}\ }\textbf
  {\bibinfo {volume} {7}},\ \bibinfo {pages} {1} (\bibinfo {year}
  {2013})}\BibitemShut {NoStop}%
\bibitem [{\citenamefont {Chung}\ \emph {et~al.}(2018)\citenamefont {Chung},
  \citenamefont {Paranjape}, \citenamefont {Dames}, \citenamefont {Shen},\ and\
  \citenamefont {Kumar}}]{Chung-PDSK-2018tor}%
  \BibitemOpen
  \bibfield  {author} {\bibinfo {author} {\bibfnamefont {S.-J.}\ \bibnamefont
  {Chung}}, \bibinfo {author} {\bibfnamefont {A.~A.}\ \bibnamefont
  {Paranjape}}, \bibinfo {author} {\bibfnamefont {P.}~\bibnamefont {Dames}},
  \bibinfo {author} {\bibfnamefont {S.}~\bibnamefont {Shen}},\ and\ \bibinfo
  {author} {\bibfnamefont {V.}~\bibnamefont {Kumar}},\ }\bibfield  {title}
  {\bibinfo {title} {A survey on aerial swarm robotics},\ }\href
  {https://doi.org/10.1109/TRO.2018.2857475} {\bibfield  {journal} {\bibinfo
  {journal} {IEEE Transactions on Robotics}\ }\textbf {\bibinfo {volume}
  {34}},\ \bibinfo {pages} {837} (\bibinfo {year} {2018})}\BibitemShut
  {NoStop}%
\bibitem [{\citenamefont {Rold{\'a}n}\ \emph {et~al.}(2018)\citenamefont
  {Rold{\'a}n}, \citenamefont {Cerro},\ and\ \citenamefont
  {Barrientos}}]{Roldan-CB-2018icirs}%
  \BibitemOpen
  \bibfield  {author} {\bibinfo {author} {\bibfnamefont {J.~J.}\ \bibnamefont
  {Rold{\'a}n}}, \bibinfo {author} {\bibfnamefont {J.~D.}\ \bibnamefont
  {Cerro}},\ and\ \bibinfo {author} {\bibfnamefont {A.}~\bibnamefont
  {Barrientos}},\ }\bibfield  {title} {\bibinfo {title} {Should we compete or
  should we cooperate? applying game theory to task allocation in drone
  swarms},\ }in\ \href {https://doi.org/10.1109/IROS.2018.8594145} {\emph
  {\bibinfo {booktitle} {2018 IEEE/RSJ International Conference on Intelligent
  Robots and Systems (IROS)}}}\ (\bibinfo  {publisher} {IEEE},\ \bibinfo {year}
  {2018})\ pp.\ \bibinfo {pages} {5366--5371}\BibitemShut {NoStop}%
\bibitem [{\citenamefont {Hayat}\ \emph {et~al.}(2016)\citenamefont {Hayat},
  \citenamefont {Yanmaz},\ and\ \citenamefont {Muzaffar}}]{Hayat-YM-2016cst}%
  \BibitemOpen
  \bibfield  {author} {\bibinfo {author} {\bibfnamefont {S.}~\bibnamefont
  {Hayat}}, \bibinfo {author} {\bibfnamefont {E.}~\bibnamefont {Yanmaz}},\ and\
  \bibinfo {author} {\bibfnamefont {R.}~\bibnamefont {Muzaffar}},\ }\bibfield
  {title} {\bibinfo {title} {Survey on unmanned aerial vehicle networks for
  civil applications: A communications viewpoint},\ }\href
  {https://doi.org/10.1109/COMST.2016.2560343} {\bibfield  {journal} {\bibinfo
  {journal} {IEEE Communications Surveys and Tutorials}\ }\textbf {\bibinfo
  {volume} {18}},\ \bibinfo {pages} {2624} (\bibinfo {year}
  {2016})}\BibitemShut {NoStop}%
\bibitem [{\citenamefont {Arafat}\ and\ \citenamefont
  {Moh}(2019)}]{Arafat-M-2019itj}%
  \BibitemOpen
  \bibfield  {author} {\bibinfo {author} {\bibfnamefont {M.~Y.}\ \bibnamefont
  {Arafat}}\ and\ \bibinfo {author} {\bibfnamefont {S.}~\bibnamefont {Moh}},\
  }\bibfield  {title} {\bibinfo {title} {Localization and clustering based on
  swarm intelligence in uav networks for emergency communications},\ }\href
  {https://doi.org/10.1109/JIOT.2019.2925567} {\bibfield  {journal} {\bibinfo
  {journal} {IEEE Internet of Things Journal}\ }\textbf {\bibinfo {volume}
  {6}},\ \bibinfo {pages} {8958} (\bibinfo {year} {2019})}\BibitemShut
  {NoStop}%
\bibitem [{\citenamefont {Maxa}\ \emph {et~al.}(2017)\citenamefont {Maxa},
  \citenamefont {Mahmoud},\ and\ \citenamefont {Larrieu}}]{Maxa-ML-2017ahswm}%
  \BibitemOpen
  \bibfield  {author} {\bibinfo {author} {\bibfnamefont {J.-A.}\ \bibnamefont
  {Maxa}}, \bibinfo {author} {\bibfnamefont {M.~S.~B.}\ \bibnamefont
  {Mahmoud}},\ and\ \bibinfo {author} {\bibfnamefont {N.}~\bibnamefont
  {Larrieu}},\ }\bibfield  {title} {\bibinfo {title} {Survey on uaanet routing
  protocols and network security challenges},\ }\href@noop {} {\bibfield
  {journal} {\bibinfo  {journal} {Adhoc and Sensor Wireless Networks}\ }\textbf
  {\bibinfo {volume} {37}} (\bibinfo {year} {2017})}\BibitemShut {NoStop}%
\bibitem [{\citenamefont {Mairaj}\ \emph {et~al.}(2019)\citenamefont {Mairaj},
  \citenamefont {Baba},\ and\ \citenamefont {Javaid}}]{Mairaja-BJ-2019smpt}%
  \BibitemOpen
  \bibfield  {author} {\bibinfo {author} {\bibfnamefont {A.}~\bibnamefont
  {Mairaj}}, \bibinfo {author} {\bibfnamefont {A.~I.}\ \bibnamefont {Baba}},\
  and\ \bibinfo {author} {\bibfnamefont {A.~Y.}\ \bibnamefont {Javaid}},\
  }\bibfield  {title} {\bibinfo {title} {Application specific drone simulators:
  Recent advances and challenges},\ }\href
  {https://doi.org/j.simpat.2019.01.004} {\bibfield  {journal} {\bibinfo
  {journal} {Simulation Modelling Practice and Theory}\ }\textbf {\bibinfo
  {volume} {94}},\ \bibinfo {pages} {100} (\bibinfo {year} {2019})}\BibitemShut
  {NoStop}%
\bibitem [{\citenamefont {V{\'a}s{\'a}rhelyi}\ \emph
  {et~al.}(2018)\citenamefont {V{\'a}s{\'a}rhelyi}, \citenamefont {Vir{\'a}gh},
  \citenamefont {Somorjai}, \citenamefont {Nepusz}, \citenamefont {Eiben},\
  and\ \citenamefont {Vicsek}}]{Vasarhelyi-VSNEV-2018sr}%
  \BibitemOpen
  \bibfield  {author} {\bibinfo {author} {\bibfnamefont {G.}~\bibnamefont
  {V{\'a}s{\'a}rhelyi}}, \bibinfo {author} {\bibfnamefont {C.}~\bibnamefont
  {Vir{\'a}gh}}, \bibinfo {author} {\bibfnamefont {G.}~\bibnamefont
  {Somorjai}}, \bibinfo {author} {\bibfnamefont {T.}~\bibnamefont {Nepusz}},
  \bibinfo {author} {\bibfnamefont {A.~E.}\ \bibnamefont {Eiben}},\ and\
  \bibinfo {author} {\bibfnamefont {T.}~\bibnamefont {Vicsek}},\ }\bibfield
  {title} {\bibinfo {title} {Optimized flocking of autonomous drones in
  confined environments},\ }\href {https://doi.org/10.1126/scirobotics.aat3536}
  {\bibfield  {journal} {\bibinfo  {journal} {Science Robotics}\ }\textbf
  {\bibinfo {volume} {3}},\ \bibinfo {pages} {3536} (\bibinfo {year}
  {2018})}\BibitemShut {NoStop}%
\bibitem [{\citenamefont {Albani}\ \emph {et~al.}(2022)\citenamefont {Albani},
  \citenamefont {Manoni}, \citenamefont {Saska},\ and\ \citenamefont
  {Ferrante}}]{Albani-MSF-2022icra}%
  \BibitemOpen
  \bibfield  {author} {\bibinfo {author} {\bibfnamefont {D.}~\bibnamefont
  {Albani}}, \bibinfo {author} {\bibfnamefont {T.}~\bibnamefont {Manoni}},
  \bibinfo {author} {\bibfnamefont {M.}~\bibnamefont {Saska}},\ and\ \bibinfo
  {author} {\bibfnamefont {E.}~\bibnamefont {Ferrante}},\ }\bibfield  {title}
  {\bibinfo {title} {Distributed three dimensional flocking of autonomous
  drones},\ }\href {https://doi.org/10.1109/ICRA46639.2022.9811633} {\bibfield
  {journal} {\bibinfo  {journal} {2022 International Conference on Robotics and
  Automation (ICRA)}\ ,\ \bibinfo {pages} {6904}} (\bibinfo {year}
  {2022})}\BibitemShut {NoStop}%
\bibitem [{\citenamefont {Chen}\ \emph {et~al.}(2020)\citenamefont {Chen},
  \citenamefont {Liu},\ and\ \citenamefont {Guo}}]{Chen-LG-2020tvt}%
  \BibitemOpen
  \bibfield  {author} {\bibinfo {author} {\bibfnamefont {W.}~\bibnamefont
  {Chen}}, \bibinfo {author} {\bibfnamefont {J.}~\bibnamefont {Liu}},\ and\
  \bibinfo {author} {\bibfnamefont {H.}~\bibnamefont {Guo}},\ }\bibfield
  {title} {\bibinfo {title} {Achieving robust and efficient consensus for
  large-scale drone swarm},\ }\href {https://doi.org/10.1109/TVT.2020.3036833}
  {\bibfield  {journal} {\bibinfo  {journal} {IEEE Transactions on Vehicular
  Technology}\ }\textbf {\bibinfo {volume} {69}},\ \bibinfo {pages} {15867 }
  (\bibinfo {year} {2020})}\BibitemShut {NoStop}%
\bibitem [{\citenamefont {Han}\ \emph {et~al.}(2007)\citenamefont {Han},
  \citenamefont {Rossi},\ and\ \citenamefont {Shen}}]{Han-RS-2007robo}%
  \BibitemOpen
  \bibfield  {author} {\bibinfo {author} {\bibfnamefont {X.}~\bibnamefont
  {Han}}, \bibinfo {author} {\bibfnamefont {L.~F.}\ \bibnamefont {Rossi}},\
  and\ \bibinfo {author} {\bibfnamefont {C.-C.}\ \bibnamefont {Shen}},\
  }\bibfield  {title} {\bibinfo {title} {Autonomous navigation of wireless
  robot swarms with covert leaders},\ }in\ \href@noop {} {\emph {\bibinfo
  {booktitle} {Proceedings of the 1st International Conference on Robot
  Communication and Coordination}}},\ Vol.~\bibinfo {volume} {8}\ (\bibinfo
  {publisher} {IEEE Press},\ \bibinfo {year} {2007})\ pp.\ \bibinfo {pages}
  {1--8}\BibitemShut {NoStop}%
\bibitem [{\citenamefont {Jain}\ \emph {et~al.}(2003)\citenamefont {Jain},
  \citenamefont {Simsek},\ and\ \citenamefont {Varaiya}}]{Jain-SV-2003cdc}%
  \BibitemOpen
  \bibfield  {author} {\bibinfo {author} {\bibfnamefont {R.}~\bibnamefont
  {Jain}}, \bibinfo {author} {\bibfnamefont {T.}~\bibnamefont {Simsek}},\ and\
  \bibinfo {author} {\bibfnamefont {P.}~\bibnamefont {Varaiya}},\ }\bibfield
  {title} {\bibinfo {title} {Control under communication constraints},\
  }\bibfield  {journal} {\bibinfo  {journal} {Proceedings of the 41st IEEE
  Conference on Decision and Control, 2002.}\ }\href
  {https://doi.org/10.1109/CDC.2002.1184366} {10.1109/CDC.2002.1184366}
  (\bibinfo {year} {2003}),\ \bibinfo {note} {date of Conference: 10-13
  December 2002; Las Vegas, NV, USA}\BibitemShut {NoStop}%
\bibitem [{\citenamefont {Tatikonda}\ and\ \citenamefont
  {Mitter}(2004)}]{Tatikonda-M-2004tac}%
  \BibitemOpen
  \bibfield  {author} {\bibinfo {author} {\bibfnamefont {S.}~\bibnamefont
  {Tatikonda}}\ and\ \bibinfo {author} {\bibfnamefont {S.}~\bibnamefont
  {Mitter}},\ }\bibfield  {title} {\bibinfo {title} {Control under
  communication constraints},\ }\href {https://doi.org/10.1109/TAC.2004.831187}
  {\bibfield  {journal} {\bibinfo  {journal} {IEE Transactions on Automatic
  Control}\ }\textbf {\bibinfo {volume} {49}},\ \bibinfo {pages} {1056}
  (\bibinfo {year} {2004})}\BibitemShut {NoStop}%
\bibitem [{\citenamefont {Claridge}\ \emph {et~al.}(2022)\citenamefont
  {Claridge}, \citenamefont {Todd}, \citenamefont {Kinsler}, \citenamefont
  {Elliott}, \citenamefont {Holman}, \citenamefont {Mitchell},\ and\
  \citenamefont {Wilson}}]{CEME-DASA-2022}%
  \BibitemOpen
  \bibfield  {author} {\bibinfo {author} {\bibfnamefont {R.}~\bibnamefont
  {Claridge}}, \bibinfo {author} {\bibfnamefont {A.}~\bibnamefont {Todd}},
  \bibinfo {author} {\bibfnamefont {P.}~\bibnamefont {Kinsler}}, \bibinfo
  {author} {\bibfnamefont {A.}~\bibnamefont {Elliott}}, \bibinfo {author}
  {\bibfnamefont {S.}~\bibnamefont {Holman}}, \bibinfo {author} {\bibfnamefont
  {C.}~\bibnamefont {Mitchell}},\ and\ \bibinfo {author} {\bibfnamefont
  {R.~E.}\ \bibnamefont {Wilson}},\ }\href@noop {} {\emph {\bibinfo {title}
  {Autonomous Reconnection of Swarming Drones within an Adversarial EM
  Environment}}},\ \bibinfo {type} {Tech. Rep.}\ (\bibinfo {year}
  {2022})\BibitemShut {NoStop}%
\bibitem [{\citenamefont {Geckil}\ and\ \citenamefont
  {Anderson}(2009)}]{GeckilAnderson-AGAME}%
  \BibitemOpen
  \bibfield  {author} {\bibinfo {author} {\bibfnamefont {I.~K.}\ \bibnamefont
  {Geckil}}\ and\ \bibinfo {author} {\bibfnamefont {P.~L.}\ \bibnamefont
  {Anderson}},\ }\href@noop {} {\emph {\bibinfo {title} {Applied Game Theory
  and Strategic Behavior}}}\ (\bibinfo  {publisher} {Chapman and Hall},\
  \bibinfo {year} {2009})\ p.\ \bibinfo {pages} {230}\BibitemShut {NoStop}%
\bibitem [{\citenamefont {Farooqui}\ and\ \citenamefont
  {Niazi}(2016)}]{Farooqui-N-2016casm}%
  \BibitemOpen
  \bibfield  {author} {\bibinfo {author} {\bibfnamefont {A.~D.}\ \bibnamefont
  {Farooqui}}\ and\ \bibinfo {author} {\bibfnamefont {M.~A.}\ \bibnamefont
  {Niazi}},\ }\bibfield  {title} {\bibinfo {title} {Game theory models for
  communication between agents: a review},\ }\href
  {https://doi.org/10.1186/s40294-016-0026-7} {\bibfield  {journal} {\bibinfo
  {journal} {Complex Adaptive Systems Modeling}\ }\textbf {\bibinfo {volume}
  {4}},\ \bibinfo {pages} {13} (\bibinfo {year} {2016})}\BibitemShut {NoStop}%
\bibitem [{\citenamefont {Diekert}(2012)}]{Diekert-2012s}%
  \BibitemOpen
  \bibfield  {author} {\bibinfo {author} {\bibfnamefont {F.~K.}\ \bibnamefont
  {Diekert}},\ }\bibfield  {title} {\bibinfo {title} {The tragedy of the
  commons from a game-theoretic perspective},\ }\href
  {https://doi.org/10.3390/su4081776} {\bibfield  {journal} {\bibinfo
  {journal} {Sustainability}\ }\textbf {\bibinfo {volume} {4}},\ \bibinfo
  {pages} {1776} (\bibinfo {year} {2012})}\BibitemShut {NoStop}%
\bibitem [{\citenamefont {Coscia}(2021)}]{Coscia-NetworkAtlas}%
  \BibitemOpen
  \bibfield  {author} {\bibinfo {author} {\bibfnamefont {M.}~\bibnamefont
  {Coscia}},\ }\bibfield  {title} {\bibinfo {title} {The atlas for the aspiring
  network scientist},\ }\href {https://arxiv.org/abs/2101.00863} {\bibfield
  {journal} {\bibinfo  {journal} {Arxiv}\ } (\bibinfo {year} {2021})},\ \Eprint
  {https://arxiv.org/abs/2101.00863} {2101.00863} \BibitemShut {NoStop}%
\bibitem [{\citenamefont {Kinsler}\ \emph {et~al.}(2022)\citenamefont
  {Kinsler}, \citenamefont {Elliott}, \citenamefont {Holman}, \citenamefont
  {Mitchell},\ and\ \citenamefont {Wilson}}]{Kinsler-EHMW-2022anglen}%
  \BibitemOpen
  \bibfield  {author} {\bibinfo {author} {\bibfnamefont {P.}~\bibnamefont
  {Kinsler}}, \bibinfo {author} {\bibfnamefont {A.}~\bibnamefont {Elliott}},
  \bibinfo {author} {\bibfnamefont {S.}~\bibnamefont {Holman}}, \bibinfo
  {author} {\bibfnamefont {C.}~\bibnamefont {Mitchell}},\ and\ \bibinfo
  {author} {\bibfnamefont {R.~E.}\ \bibnamefont {Wilson}},\ }\bibfield  {title}
  {\bibinfo {title} {Broadcast vs narrowcast: risk and reconnection time in a
  swarm of stealthy agents},\ }\href@noop {} {\bibfield  {journal} {\bibinfo
  {journal} {Draft}\ } (\bibinfo {year} {2022})}\BibitemShut {NoStop}%
\end{thebibliography}
